%% file: main.tex
\documentclass[sigconf,natbib=false,nonacm]{acmart}
\usepackage{flushend}
\usepackage{caption}
\usepackage{subcaption}
\usepackage{tabularx}
\usepackage{booktabs} 
\usepackage{xcolor}
\usepackage{tikz}
\usepackage{graphicx}
\usepackage{colortbl}
\usepackage{placeins}
\usepackage{booktabs}
\usepackage{cleveref}
\usepackage{hyperref}
\usepackage{appendix}
\usepackage{makecell}

\crefname{figure}{Fig.}{Fig.}
\Crefname{figure}{Fig.}{Fig.}

\crefname{table}{Tab.}{Tab.}
\Crefname{table}{Tab.}{Tab.}

\crefname{section}{Sec.}{Sec.}
\Crefname{section}{Sec.}{Sec.}

\usetikzlibrary{automata, positioning, chains, shapes.symbols, shapes.geometric}

\usepackage{textcomp}

\AtBeginDocument{%
  }

\setcopyright{acmcopyright}
\copyrightyear{2025}
\acmYear{2025}
\acmDOI{XXXXXXX.XXXXXXX}

\RequirePackage[
  datamodel=acmdatamodel,
  style=acmnumeric,
  natbib
  ]{biblatex}

\bibliography{bibliography}

\usepackage{todonotes}

\makeatletter
\define@key{todonotes}{Ken}[]{%
    \setkeys{todonotes}{author=Ken,color=blue}}%
\define@key{todonotes}{Markus}[]{%
    \setkeys{todonotes}{author=Markus,color=red}}%
\define@key{todonotes}{Jenny}[]{%
    \setkeys{todonotes}{author=Jenny,color=orange}}%
\define@key{todonotes}{Tim}[]{%
    \setkeys{todonotes}{author=Tim,color=green}}%
\define@key{todonotes}{Lauren}[]{%
    \setkeys{todonotes}{author=Lauren,color=purple}%
}%
\makeatother   

\definecolor{uhhblue}{rgb}{0.08,0.443,0.733}
\definecolor{uhhgreen}{rgb}{0.08,0.733,0.443}
\definecolor{Gray}{gray}{0.85}
\definecolor{LightCyan}{rgb}{0.88,1,1}

\newcolumntype{a}{>{\columncolor{Gray}}c}
\newcolumntype{b}{>{\columncolor{white}}c}

\newcolumntype{u}{>{\columncolor{uhhgreen}}l}
\newcolumntype{v}{>{\columncolor{uhhblue}}l}

\begin{document}

\title{A Hands-free Spatial Selection and Interaction Technique using Gaze and Blink Input with Blink Prediction for Extended Reality}


\author{Tim Rolff}
\authornote{Both authors contributed equally to this research.\\\\
\textcopyright 2025 Copyright held by the owner/author(s).
This is the Author’s Original Manuscript (AOM/preprint) version of the work. It is posted here for your personal use. Not for redistribution.}
\email{tim.rolff@uni-hamburg.de}
\orcid{0000-0001-9038-3196}
\affiliation{%
  \institution{University of Hamburg}
  \city{Hamburg}
  \country{Germany}
}

\author{Jenny Gabel}
\authornotemark[1]
\email{jenny.gabel@uni-hamburg.de}
\orcid{0000-0003-0180-1830}
\affiliation{%
  \institution{University of Hamburg}
  \city{Hamburg}
  \country{Germany}
}

\author{Lauren Zerbin}
\email{lzerb@post.au.dk}
\orcid{0009-0009-4681-6658}
\affiliation{%
  \institution{University of Aarhus}
  \city{Aarhus}
  \country{Denmark}
}

\author{Niklas Hypki}
\email{niklas.hypki@uni-muenster.de}
\affiliation{%
  \institution{University of Münster}
  \city{Münster}
  \country{Germany}
}

\author{Susanne Schmidt}
\email{susanne.schmidt@canterbury.ac.nz} 
\orcid{0000-0002-8162-7694}
\affiliation{%
  \institution{University of Canterbury}
  \city{Christchurch}
  \country{New Zealand}
}

\author{Markus Lappe}
\email{mlappe@uni-muenster.de}
\orcid{0000-0001-8814-7098}
\affiliation{%
  \institution{University of Münster}
  \city{Münster}
  \country{Germany}
}

\author{Frank Steinicke}
\email{frank.steinicke@uni-hamburg.de}
\orcid{0000-0001-9879-7414}
\affiliation{%
  \institution{University of Hamburg}
  \city{Hamburg}
  \country{Germany}
}

\renewcommand{\shortauthors}{Rolff\&Gabel et al.}

\begin{abstract}
Gaze-based interaction techniques have created significant interest in the field of spatial interaction. Many of these methods require additional input modalities, such as hand gestures (e.g., gaze coupled with pinch). Those can be uncomfortable and difficult to perform in public or limited spaces and pose challenges for users who are unable to execute pinch gestures.
To address these aspects, we propose a novel, hands-free \emph{Gaze+Blink} interaction technique that leverages the user's gaze and intentional eye blinks. 
This technique enables users to perform selections by executing intentional blinks. It facilitates continuous interactions, such as scrolling or drag-and-drop, through eye blinks coupled with head movements. So far, this concept has not been explored for hands-free spatial interaction techniques. 
We evaluated the performance and user experience (UX) of our \emph{Gaze+Blink} method with two user studies and compared it with \emph{Gaze+Pinch} in a realistic user interface setup featuring common menu interaction tasks. 
Study 1 demonstrated that while \emph{Gaze+Blink} achieved comparable selection speeds, it was prone to accidental selections resulting from unintentional blinks. In Study 2 we explored an enhanced technique employing a deep learning algorithms for filtering out unintentional blinks.\\ 

\end{abstract}


\keywords{interaction techniques, hand tracking, eye tracking, blink detection, human-computer interaction, mixed reality, deep-learning, blink prediction, extended reality}


\maketitle

\section{Introduction}
\begin{figure*}[t]
    \centering
    \includegraphics[width=\textwidth]{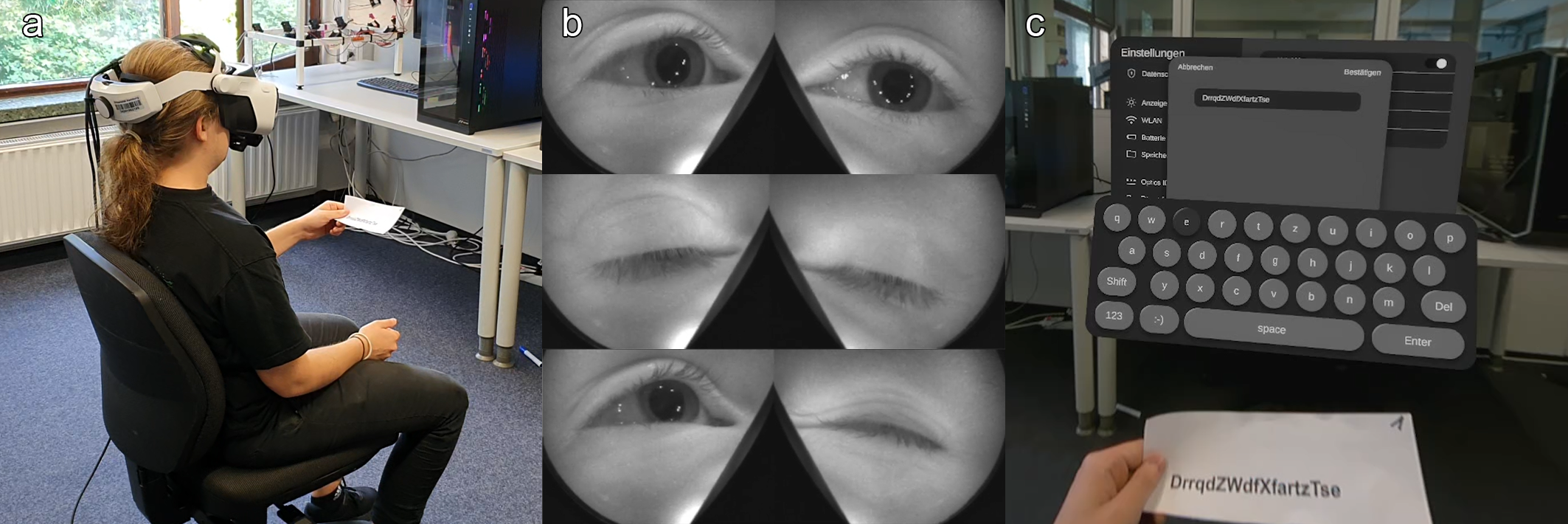}
    \caption{Gaze+Blink interaction technique: (a) user wearing the Varjo XR-4 HMD, (b) exemplary eye tracking of a blink and left eye wink, (c) spatial UI with keyboard input task used in the user study.}
    \label{fig:teaser}
\end{figure*}
\noindent Modern interaction with spatial user interfaces (UIs) requires various discrete and continuous inputs such as button selections, text input, drag-and-drop, and scrolling. Thus, these functionalities collectively define a comprehensive set of required interactions that an interaction paradigm needs to support. One emerging state-of-the-art technique is \emph{Gaze+Pinch}~\citep{Pfeuffer2017}.
This is an essential interaction technique for certain modern head-mounted displays (HMDs), such as the Apple Vision Pro (AVP) or the Microsoft HoloLens~2. 
A relevant concern of most gaze-hand interaction techniques is their visibility. While the actions performed inside the virtual environment are not immediately visible to outsiders, it is still apparent to others, that an action is performed. This may be uncomfortable 
to some users, especially in public spaces. Moreover, in constrained spaces, it might not be feasible (e.g., during flights), indiscreet, or inappropriate to perform arm movements and hand gestures. Different movement restrictions may prevent users from using hand-based interaction techniques, especially when performing continuous interactions like drag-and-drop or scrolling, where arm movements are required \cite{Pfeuffer2017}.\\

Nowadays, there is increased interest in gaze-based interaction techniques with the recent advancements of eye-tracking technology for modern commercially available HMDs, such as the Meta Quest Pro, the AVP, and the Varjo XR-4. Here, hands-free interaction techniques, such as \emph{Gaze+Dwell} \cite{Hansen03,Mott17,Mutasim2021} and \emph{Outline Pursuits}~\cite{Sidenmark2020}, provide individual selection interactions that rely on the user's gaze. 
Due to the required dwell threshold for selection confirmation, \emph{Gaze+Dwell} slows down users~\citep{Mutasim2021}. Similarly, the tracing of \emph{Outline Pursuits} requires some amount of time to perform the action.

To mitigate this time requirement of previous works, we aim to use blink as a faster and more efficient hands-free gaze-based input technique, as a blink only requires approximately $120\pm 2$ ms~\cite{duchowski2017eye}.
Various blink-based interaction techniques have been proposed in the context of gaming, keyboard input, and as an accessibility feature~\citep{Velloso16, Colley24, Lu20, Attiah21, Porter23, Rebsdorf23, Xiao19}. Using blinks for selection confirmations can be more effective compared to \emph{Gaze+Dwell}~\cite{Porter23, Lu20}.\\

Particularly interesting are the works of \citet{Gomez21, Nukarinen2016}, who previously demonstrated interactions based on various states of eye openness \cite{Gomez21} and introduced the concept of combining gaze with head rotation for system control \cite{Nukarinen2016}.
Both of these works focus on two-dimensional desktop environments. 
However, combined gaze- and blink-based continuous spatial interactions in XR has not received much attention in previous works~\cite{monteiro2021hands}. Building on these ideas, we explore a solution for discrete and continuous interactions by integrating blink- with head-based input for XR.\\

We aim to provide a feasible hands-free alternative to existing techniques 
that provides comparable performance. 
Keeping in mind the practical applicability, this technique can be implemented on any modern HMD that supports eye-tracking.  
In addition, \emph{Gaze+Blink} could offer more comfort and flexibility for interactions in public or constrained spaces. This technique could also increase the accessibility of XR devices for users with limited upper limb movement or motor control.
For this novel technique, we will examine the following research questions:
\begin{itemize}
     \item[\textbf{RQ1}] How can an interaction model be implemented that covers both discrete and continuous inputs through \emph{Gaze+Blink}?\\
     \item[\textbf{RQ2}] How does \emph{Gaze+Blink} compare to \emph{Gaze+Pinch} for different discrete and continuous UI interactions?\\
     \item[\textbf{RQ3}] What are the advantages and challenges of \emph{Gaze+Blink}?
\end{itemize}
\noindent To answer these research questions, we created an implementation (\textbf{RQ1}) and conducted two studies (\textbf{RQ2}, \textbf{RQ3}). At first, we evaluated our interaction technique against \emph{Gaze+Pinch} in a realistic user interface featuring familiar UI layouts and elements. We found that \emph{Gaze+Blink} performed similarly in terms of UX and speed, but users triggered unintended selection inputs through involuntary blinks.
Based on these results, we formulated another research question to identify potential improvements:
\begin{itemize}
     \item [\textbf{RQ4}] Are there optimization methods to reduce involuntary blinks for \emph{Gaze+Blink} that could improve this interaction technique?
\end{itemize}

To address \textbf{RQ4}, we created a deep-learning model. This model is used to classify voluntary and involuntary blinks, avoiding unintended interactions. Our model successfully detected 75\% of the involuntary blinks on uncalibrated users.
We conducted a second study, comprised of two parts: (i) We collected data to label accidental blinks and trained a deep-learning model for classifying accidental blinks. (ii) We classify involuntary blinks through our learned model, extending the blink-based interaction, creating \emph{Gaze+BlinkPlus}. During our second user study, we evaluated \emph{Gaze+Pinch}, \emph{Gaze+Blink}, and our improved method \emph{Gaze+BlinkPlus}.\\
%
%
%
%

\noindent To summarize, with our work we contribute the following:
\begin{enumerate}
    \item \emph{Gaze+Blink}, a novel hands-free, discrete, and continuous interaction technique.
    \item A machine learning challenge and solution for the classification of voluntary and involuntary eye blinks in the novel context of blink-based spatial interactions. 
    \item Two comprehensive user studies with realistic menu UIs and interaction tasks, comparing \emph{Gaze+Pinch} with our proposed \emph{Gaze+Blink} and \emph{Gaze+BlinkPlus}. 
    \item Thorough insights into interaction performance, perceived workload, qualitative aspects, and challenges for integrating blinks in interaction techniques.
\end{enumerate}

\section{Related Work}
\label{sec:Related Work}


\subsection{Gaze-Based Interaction Techniques}
As HMDs with built-in eye-tracking become increasingly available, gaze-based techniques have gained more relevance for interacting with virtual and mixed reality environments. Recently, various multimodal gaze-based input techniques have been developed and evaluated~\cite{Plopski2022EyeXR}. In this section, we will briefly overview the current state of research on relevant eye-tracking-based techniques: 

\subsubsection{Gaze \& Dwell}
The \emph{Gaze \& Dwell} technique \citep{Hansen03,Mott17,Hansen18} uses gaze for targeting and gaze focus, with a specified dwell time for target selection confirmation. This hands-free technique only requires gaze, with no additional input devices. The technique is easy to understand, but the selection is slower compared to using a controller button click or a pinch gesture~\cite{Mutasim2021}. There is a tradeoff between lowering the dwell threshold (faster dwell activation times) and an increase in error rate due to accidental selections~\citep{Mutasim2021, Mott17}. This is not ideal in standard use cases with high-frequency input, such as prolonged system UI interactions and keyboard input.
\emph{Eye\&Head Dwell}~\citep{Sidenmark19} can improve efficiency by modulating the dwell timer with head motion. 
One major drawback of basic \emph{Gaze \& Dwell} is that it does not inherently support complex interactions and continuous input (e.g., dragging or scrolling).

\subsubsection{Gaze+Pinch}
\emph{Gaze+Pinch} uses gaze for targeting and hand-tracking with a pinch gesture for selection~\citep{Pfeuffer2017,MicrosoftGaze}. It can be implemented on several modern HMDs such as the Meta Quest Pro, Microsoft HoloLens 2, AVP, and Varjo XR-4. On the HoloLens 2 and the AVP, \emph{Gaze+Pinch} is one of the standard interaction techniques. 
\citet{Mutasim2021} found no significant performance differences between pinch versus button clicks for selection confirmation. This technique facilitates fast targeting and selection of remote objects and requires less physical effort compared to techniques that involve arm movement for targeting (e.g., handray interaction) \citep{Lystbaek2022, Wagner2023}. 
Moreover,~\citet{Lystbaek2024HandsOn} found that the indirect input also requires less physical effort than performing direct gestures for object manipulation. 
However, \emph{Gaze+Pinch} also faces some challenges~\citep{Pfeuffer2024}, one of which is \emph{(Un)Learning}. Using \emph{Gaze+Pinch} initially requires rethinking of familiar behaviors based on real-world experiences. As users tend to intuitively move towards a target that they want to acquire, 
Using this technique might initially require unlearning moving the arm and hand during interaction~\citep{Pfeuffer2024}.
\emph{Early and Late Triggers} are another concern, as a correctly matched timing between the gaze targeting and gesture confirmation is required for successful input. 

\subsubsection{Gaze-Hand Alignment}
Another approach to gaze-based interaction is \emph{Gaze-Hand Alignment}~(GHA). With GHA, a selection is triggered by aligning the gaze with the fingertip or the end of a handray on the image plane~\citep{Lystbaek2022,Wagner2023}. While GHA outperforms classic handray selection techniques for mid-sized and large targets, users are required to use their hand or finger for the selection trigger alignment, adding more physical motion to the interaction~\cite{Wagner2023,Aziz24}.

\subsubsection{Gaze-assisted Pointing Techniques}
Various gaze-assisted pointing techniques have been proposed to facilitate the selection of small and distant targets and reduce selection induced errors. \emph{RayToGaze}~\citep{gabel2023redirecting} leverages gaze to subtly redirect hand-rays without altering the core pointing interaction. A proximity-based redirection is applied that gradually pulls the ray towards the center of the gaze position. This increases selection performance compared to classic handray. 
\emph{MAGIC pointing} is a fundamental technique that warps the cursor to the gaze location, where the selection can then be adjusted by manual input~\citep{Fares13, Zhai99}. The MAGIC approach significantly increases pointing speed. \emph{EyePointing} extends the \emph{MAGIC} concept by using gaze for targeting and  mid-air pointing for confirmation of remote selections~\citep{Schweigert2019}.
\emph{Outline Pursuits}~\citep{Sidenmark2020} matches the user's smooth pursuit eye movement to a moving stimulus (outline) to confirm selection of partially occluded targets. \emph{GazeRayCursor}~\citep{Chen23} uses gaze for target depth estimation and highlights the closest target to resolve ambiguities for raycast selection.

\subsubsection{Blink-Controlled Interaction Techniques:} \label{sec:blink-selection}
Voluntary blinking with both eyes or winking with one eye can be used for hands-free interaction. ~\citet{Gomez21} proposed \emph{Gaze+Hold}, an eyes-only interaction technique with different interaction modes. Gaze is used for target indication and the closing of one eye indicates input, while the movement of the open eye provides continuous input. They found that \emph{Gaze+Hold} is effective and there were no significant differences in performance, usability, or workload between the use of the dominant versus non-dominant eye.
Blinking and winking have further been proposed as alternatives to mouse input in the contexts of gaming~\citep{Velloso16, Colley24}, as an accessibility feature~\citep{Missimer10}, and for virtual keyboard input~\citep{Lu20, Attiah21, Porter23}. 
\citet{Lu20} as well as \citet{Porter23} found that blinking can improve hands-free input performance compared to \emph{Gaze \& Dwell}.
Recent research has also explored Blink Interaction Techniques (BITs) for VR applications \citep{Rebsdorf23}. \citet{Zenner21} used blinks to leverage change blindness~\citep{Suma11} for undetected hand redirection. \citet{Xiao19} created a VR music-on-demand system for patients with limb paralysis, where users could use blinking for searching and selecting songs. 
~\citet{jota2015palpebrae} explored the design space of different eyelid gestures in a desktop environment in combination with other input modalities, such as mouse input.
~\citet{Rebsdorf23} proposed five different BITs in a VR gaming context. Blinking and winking are used for interactions with the virtual environment, such as teleporting, shooting, and flying. The study results indicated that BITs worked reasonably well. 
However, potential fatigue due to rapid blinking and reduced accessibility of single-eyed interactions have to be considered. With the improved eye-tracking accuracy and increased availability of built-in eye-trackers in recent HMDs, blinking and eyelid gestures seem to be a promising modality for spatial interaction.

\subsubsection{Gaze-Head Interaction Techniques}\label{sec:gazeheadbased-interaction}
While existing research mainly investigated the use of head movements as a subsequent refinement for gaze-based pointing~\citep{Sidenmark19}, some research has also explored the integrated, combined use of gaze- and head-tracking for interaction.
\citet{Nukarinen2016} proposed the concept of \emph{HeadTurn}, an interaction technique that combines head rotation with gaze for system control. The gaze direction is used for target indication (selecting the control interface), and right/left head rotation is used to change the value of the associated control parameter.
They tested this interaction technique for increasing and decreasing numbers in a desktop UI and found that participants generally had a positive experience with using this technique.\\

\emph{Eye\&Head} is another concept by~\citet{Sidenmark19} that integrates head movement with gaze interaction for VR interactions. It defines three novel interaction techniques: \emph{Eye\&Head Pointing}, \emph{Eye\&Head Dwell}, and \emph{Eye\&Head Convergence} that support pointing, hovering, visual exploration around pre-selections, as well as iterative and fast confirmation for targets. They showed that \emph{Eye\&Head} provides more control and flexibility for spatial interactions.

\subsection{Blink Classification}
\label{sec: Blink Classification}
\begin{figure}[b]
\centering
    \resizebox{0.95\columnwidth}{!}{
        \input{images/states}
    }
    \caption{State graph for our discrete (blue rectangles) and continuous (green diamonds) \emph{Gaze+Blink} interactions. For more details, see. \cref{sec: Gaze+Blink Interaction}.}
    \label{fig: states}
\end{figure}
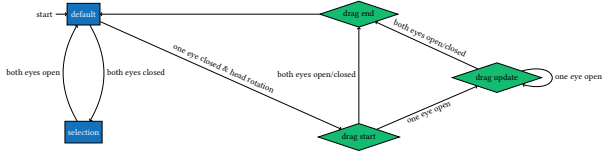
\begin{figure*}[h]
    \centering
    \includegraphics[width=\linewidth]{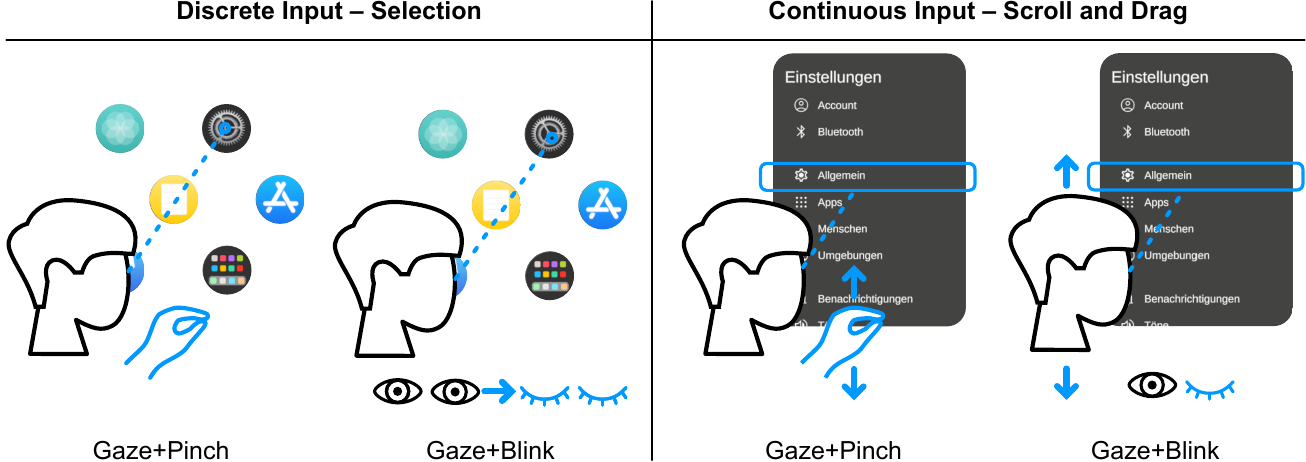} 
\caption{Comparison of the Gaze+Pinch and Gaze+Blink interaction techniques for discrete and continuous input.}
\label{fig:techniques-comparison}
\end{figure*}

The classification of different eye movements is a fundamental area of eye-tracking research, focusing on the detection and categorization of different gaze behaviors, such as saccades, fixations, and smooth pursuits \cite{startsev20191d,zemblys2019gazenet,agtzidis2016smooth,komogortsev2013automated,andersson2017one,salvucci2000identifying,dar2021remodnav,rolff2022saccades,rolff2023deep}. Even though there is a vast amount of literature on gaze event classification, the study of blink classification remains relatively underexplored. While there are some works that aim at classifying blinks in eye-tracking signals \cite{nystrom2024blink}, these do not distinguish between different types of blinks. Notably, blinks encompass two distinct types that have evolved based on their underlying mechanisms: voluntary blinks that are consciously performed during activities such as face-to-face communication \citep{volkmann1986human,homke2018eye}, and involuntary blinks, which occur subconsciously and serve as a fundamental reflex.\\

\textbf{Voluntary blinks} can be divided into two types \citep{sato2017automatic}: (i) those executed over an extended duration, and (ii) those performed firmly but briefly.\\

For \textbf{Involuntary blinks}, there are also different subclassifications \citep{zenner2023induce}. Notably, it is possible to categorize involuntary blinks as either spontaneous or reflexive blinks \citep{manning1983reflex}, with spontaneous blinks naturally occurring 6–30 times per minute in a virtual environment \citep{kim2022change}, amounting to an average of 17 blinks per minute \citep{bentivoglio1997analysis}, with a typical duration of 100-150 ms \citep{volkmann1986human}.\\

Consequently, blink duration alone is insufficient to definitively identify a blink as voluntary. Existing methodologies that are capable of differentiating between voluntary and involuntary blinks frequently rely on raw video data from the eye cameras for their predictions \citep{abe2013automatic,sato2015automatic,sato2017automatic}.  Alternatively, they depend on additional modalities unavailable on typical virtual reality devices, such as electroencephalography (EEG) \citep{agarwal2019blink,giudice20201d}. To our knowledge, no current approach exists that purely leverages the output markers of eye-trackers, such as gaze direction, eye openness, or pupil size for this classification.

\section{Gaze+Blink Interaction}
\label{sec: Gaze+Blink Interaction}
To answer \textbf{RQ1}, we utilize eye closure by combining gaze- and blink- based input modalities for discrete selections, following the example of existing works (cf. \cref{sec:blink-selection}).
As continuous interactions are more challenging, we designed an interaction method that integrates gaze-, blink-, and head-tracking modalities. We drew inspiration from previously mentioned approaches that use eye openness and head interactions (see~\cref{sec:gazeheadbased-interaction}).
To map these input requirements, we defined the following five input states for \emph{Gaze+Blink}: 

\begin{itemize}
    \item \textbf{Default:} When both eyes are open
    \item \textbf{Selection:} When both eyes are closed
    \item \textbf{Drag start:} Marked by closing of only one eye
    \item \textbf{Drag update:} When one eye is closed \& head movement 
    \item \textbf{Drag end:} When the eye is opened again
\end{itemize}

\noindent These five states are described in the state graph for our interaction mechanism (cf.~\cref{fig: states}). If both eyes are open,  
we are in the default state, performing no interaction. If both eyes are closed 
below a threshold pre-calibrated for each user, a selection confirmation is performed. From the selection state, it is only possible to go back to the default state to avoid unintended drag interactions. If only one eye is closed and head movement is detected, we perform a drag start interaction until both eyes are open or fully closed.\\

\noindent These five states 
are used to define the two input modes for \emph{Gaze+Blink}:
\begin{enumerate}
    \item[(D)] discrete input for selection
    \item[(C)] continuous input for scroll and drag
\end{enumerate}

\subsection{(D) Discrete Selection}

\begin{figure}[H]
    \centering
    \includegraphics[width=0.9\linewidth]{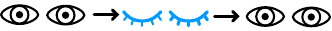} 
\caption{Gaze+Blink Selection Sequence}
\label{fig:gaze-blink-select}
\end{figure}

Figure~\ref{fig:techniques-comparison}
illustrates the discrete selection task with \emph{Gaze+Blink} and compares it with the equivalent discrete interaction in \emph{Gaze+Pinch}~\citep{Pfeuffer2017}. \emph{Gaze+Blink} selection is composed of target indication by looking at the target and selection confirmation by blinking, as viualized in Figure~\ref{fig:gaze-blink-select}:
\begin{enumerate}
    \item Default state, both eyes open $\Rightarrow$ target indication
    \item Selection state:
    \begin{enumerate}
        \item[i] start: both eyes closed $\Rightarrow$ target selection
        \item[ii] end: both eyes opened $\Rightarrow$ return to default state
    \end{enumerate}
\end{enumerate}
\noindent The blinking is detected with a specified eye openness threshold value. As a blink can be performed nearly instantly by the user ($120\pm2$ms) \cite{duchowski2017eye}, this allows for a fast update of the blink selection confirmation state.
In contrast, \emph{Gaze \& Dwell} has higher selection times and lower throughput than \emph{Gaze+Pinch}~\cite{Mutasim2021} due to the required dwell activation threshold (recommended around $300$ms)~\cite{Mutasim2021}.

\subsection{(C) Continuous Scroll and Drag}
\begin{figure}[H]
    \centering
    \includegraphics[width=0.9\linewidth,trim={0 0.25cm 0 0.25cm},clip]
    {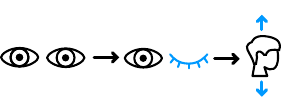} 
\caption{Gaze+Blink Scroll Sequence}
\label{fig:gaze-blink-scroll}
\end{figure}

For continuous input, we designed \emph{Gaze+Blink} for scrolling and drag-and-drop. As illustrated in Figure~\ref{fig:gaze-blink-scroll}, the movement is performed by closing one eye and moving the head into the desired direction. 
Since the movement is controlled based on the head rotation and not the gaze position, this enables users to visually inspect the movement and their surrounding environment. 
The sequence for this continuous input is defined as follows:

\begin{enumerate}
    \item Default state, both eyes open $\Rightarrow$ target indication\\
    \item Drag state:
    \begin{enumerate}
        \item[i] start: one eye closed $\Rightarrow$ target locked
        \item[ii] update: one eye closed \& head rotation\\ $\Rightarrow$ target movement
        \item[iii] end: both eyes opened $\Rightarrow$ target released, return to default state
    \end{enumerate}
\end{enumerate}

\noindent The rotation is calculated based on raycasting onto a plane, where the intersection point is then used as the origin. Depending on the delta between the points (angular movement), the UI object is then moved by the rotational gain amount. This means that we do not have a linear Control-Display (C/D) ratio but an angular mapping of the movement, similar to raycasting. Therefore, the (C/D) ratio between head-rotation and angular gain of the object movement increases with distance from the head to the target object.\\

A threshold for distinguishing between discrete and continuous input as with \emph{Gaze+Pinch} is not necessary with our technique, as this is specified through the blink state (one versus two eyes closed). This continuous input matches scrolling or drag-and-drop tasks performed with \emph{Gaze+Pinch} (cf. Figure~\ref{fig:techniques-comparison}). 

In summary, \emph{Gaze+Blink} provides analogous discrete and continuous interactions to \emph{Gaze+Pinch}. 
To our knowledge, so far, no techniques for implementing continuous interactions have been proposed for 
hands-free interactions in XR.
Existing hands-free gaze-only spatial interactions have lower performance compared to blink-based techniques~\citep{Porter23, Lu20}.
Our technique is intended for interactions with efficiently performed tasks and fast and frequent inputs. Based on these properties, we find \emph{Gaze+Pinch} to be a suitable baseline for testing our proposed approach.
A comparison of the similarities and differences between both blink-based interaction techniques is listed in~\cref{tab:comparison}.  


\iffalse
    \begin{table}[b]
    \caption{Gaze Interaction Techniques Comparison}
    \centering
    \resizebox{\columnwidth}{!}{%
    \begin{tabular}{r|l|l}
    \toprule
         & \textbf{Gaze+Pinch} & \textbf{Gaze+Blink} \\
    \midrule
    \textbf{Target Indication} & gaze & gaze \\
    \textbf{Selection Confirmation} & pinch with index and thumb & blink with both eyes\\
    \textbf{Continuous Movement} & hold pinch and drag with hand & close one eye with one eye and rotate head \\
    \textbf{Modalities} & gaze, hand & gaze, head\\
    \bottomrule
    \end{tabular}%
    }
    \label{tab:comparison}
    \end{table}
\else
    \begin{table}[b]
    \caption{Gaze Interaction Techniques Comparison}
    \centering
    \resizebox{\columnwidth}{!}{%
    \begin{tabular}{l|c|c|c|c}
    \toprule
    \textbf{Technique} & \makecell{\textbf{Target}\\\textbf{Indication}} & \makecell{\textbf{Selection}\\\textbf{Confirmation}} & \makecell{\textbf{Continuous}\\\textbf{Movement}} & \textbf{Modalities} \\
    \midrule
    \textbf{Gaze+Pinch} & gaze & \makecell{pinch with index\\and thumb} & \makecell{hold pinch and\\drag with hand} & \makecell{gaze\\hand} \\
    \cline{0-4}
    \textbf{Gaze+Blink} & gaze & blink with both eyes & \makecell{close one eye\\and rotate head} & \makecell{gaze\\head} \\
    \bottomrule
    \end{tabular}%
    }
    \label{tab:comparison}
    \end{table}
\fi

\section{Study 1}
\label{sec: First Study}
\subsection{Methodology}
\subsubsection{Study design:}
\label{sec: First Study -- Study Design}
\begin{figure*}[t]
    \resizebox{\textwidth}{!}{%
        \input{images/studyprocedure}
    }
    \caption{Sequence of interactions to solve the task that was set to the participants.}
    \label{fig: task}
\end{figure*}
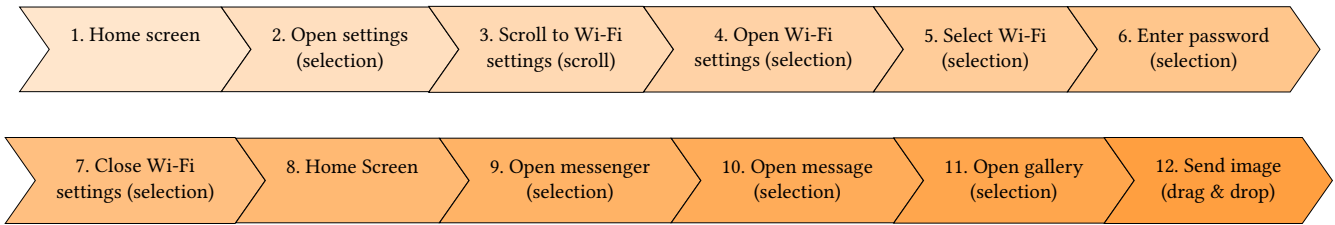
We conducted study 1 for evaluating the effectiveness and to uncover potential usability issues of our proposed interaction technique. 
The study was designed as a stationary and seated mixed reality (MR) experiment. To have a realistic comparison against a consumer-grade device that utilizes \emph{Gaze+Pinch} as its main interaction technique, we created a scenario focusing on typical UI interactions. Since the AVP is the most recent and prominent example of a \emph{Gaze+Pinch} interface, we designed an interface that is inspired by it in visual appearance, layout, and behavior.\\

For our setup, we utilized a \mbox{Varjo XR-4} (cf.~\cref{sec: Technical Implementation and Assumptions - First Study} and \cref{sec: Technical Implementation and Assumptions - Second Study}), a PC MR HMD, directly connected to a Windows computer through a tethered connection.
The study was conducted with the integrated eye-tracker of the \mbox{Varjo XR-4}. We set our render resolution to $3840\times3744$ pixels and used a field of view (FoV) of $120^\circ\times105^\circ$.
We used a desktop computer with an Intel Core i14700k CPU and an NVIDIA RTX 4080 with 32 GB of RAM. Our implementation is based on the Unity 3D\footnote{\url{unity.com/}} real-time development engine.  

\subsubsection{Study Tasks:}\label{par:studytasks}
As the focus of our initial evaluation of \emph{Gaze+Blink} was on basic UI interactions, we did not consider 3D object manipulation, including scaling and rotation.
We created commonly used discrete and continuous UI menu and system control interaction tasks:
\begin{enumerate}
    \item \textbf{Selection task:} Select the correct icons from the main menu (first Settings, later Messenger icon).
    \item \textbf{Scroll task:} Scroll through the settings menu to find Wi-Fi settings, select Wi-Fi Settings, select toggle button.
    \item \textbf{Text input task:} Select the correct keys on the virtual keyboard to input the password, click on confirm.
    \item \textbf{Drag \& drop task:} Drag an image from a folder into a messenger window.
\end{enumerate}
\noindent We designed a scenario where each participant had to send a picture through the messaging application, but needed to connect to Wi-Fi first (cf.~\cref{fig: task}). The task started on the home screen of the interface without a Wi-Fi connection. Users were then tasked to establish an internet connection by finding the settings menu on the home screen (\emph{selection task}). Afterwards, they needed to find the Wi-Fi setting in the settings menu (\emph{scroll task}) and enable the Wi-Fi connection (\emph{selection task}). Participants were then asked to insert the correct Wi-Fi password handed out on a piece of paper (\emph{text input task}) using the pass-through of the HMD. Before entering the password, participants had to read the password out loud to ensure that they could identify the letters. Once the correct password was entered, users needed to navigate back to the messaging app (\emph{selection task}) and drag a specific image into the text field (\emph{drag \& drop task}). The full sequence of interactions can also be seen in~\cref{fig: task}.\\

Based on our research questions, we aimed to investigate if there are significant differences regarding performance and UX between \emph{Gaze+Pinch} and \emph{Gaze+Blink}:
\begin{enumerate}
    \item[\emph{H1a:}] Task performance, measured as task completion time and error rate, differs significantly between \emph{Gaze+Pinch} and \emph{Gaze+Blink}.
    \item[\emph{H2a:}] Perceived workload differs significantly between \emph{Gaze+Pinch} and \emph{Gaze+Blink}.
    \item[\emph{H3a:}] UX differs significantly between \emph{Gaze+Pinch} and \emph{Gaze+Blink}.
\end{enumerate}

\subsubsection{Technical Implementation \& Assumptions:}
\label{sec: Technical Implementation and Assumptions - First Study}
\begin{figure*}[t]
    \centering
    \begin{subfigure}[t]{0.49\textwidth}
        \centering
        \includegraphics[width=\textwidth]{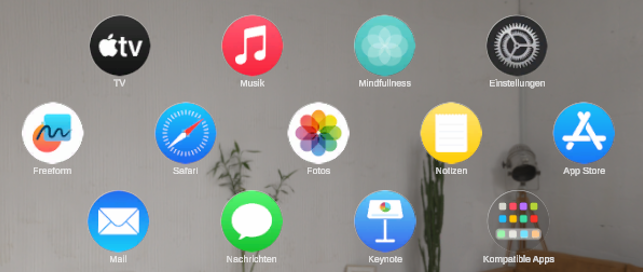}
        \caption{
            Home screen of our interface with multiple applications.}
    \end{subfigure}%
    \hspace{0.8em}%
    \begin{subfigure}[t]{0.49\textwidth}
        \centering
        \includegraphics[width=\textwidth]{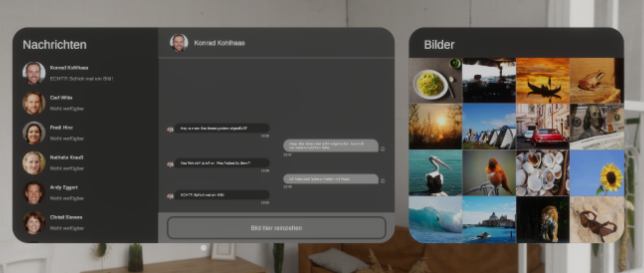}
        \caption{
            Messenger with chat history and an image gallery window.}
    \end{subfigure}
    \caption{Sample views of different menus and apps used during our study.}
    \label{fig:avp-uimenu}
\end{figure*}
Due to the privacy restriction of the AVP\footnote{\url{apple.com/privacy/docs/Apple\_Vision\_Pro\_Privacy\_Overview.pdf}} not allowing us to access the raw eye tracking position, we performed all our studies on the \mbox{Varjo XR-4} that provides similar specifications to the AVP platform\footnote{\url{varjo.com/products/xr-4/}}.
For the gaze interaction, we rely on the built-in eye-tracker of the Varjo~XR-4 to estimate the gaze position and perform a raycast for the interaction with the UI elements. To provide realistic interaction tasks involving familiar and tested UI layouts, we decided to implement our UI design based on Apple's recently released VisionOS interface. It offers a standardized and recognizable GUI menu with established UI patterns, which provide common tasks and applications, such as a messenger, settings menu, photo gallery, and a virtual keyboard, as illustrated in \cref{fig:avp-uimenu}.
We built the application in the Unity game engine from Unity Technologies, version 2022.3.3f1, and implemented the UI elements with the Nova UI Framework \footnote{\url{novaui.io/}} using VisionOS App Icons from Figma\footnote{"\href{https://www.figma.com/community/file/1304822022602387338/visionos-app-icons-free-resource}{VisionOS App Icons • Free Resource}" by \href{https://www.figma.com/@david_akh}{David Akhmedbayev}, \href{https://creativecommons.org/licenses/by/4.0/}{CC BY 4.0}}.
The UI is placed at 2.5m distance from the user according to guidelines for spatial interfaces~\citep{Bergstroem2021, MicrosoftCriteria}. 
To ensure usability and a comfortable interaction, the sizes of the UI elements were also chosen according to guidelines~\citep{Bergstroem2021, MetaUserInterface}. Overall, the angular sizes of the used UI elements are between $1.06$\textdegree{} and $22.23$\textdegree{}. A table with detailed angular sizes for each specific selectable UI element is listed in the Appendix in~\cref{tab:ui-angular}.
While the reconstruction of missing data from eye tracking systems is quite common \cite{grootjen2024uncovering}, 
we would rather not interpolate with future data to avoid additional latency, as we intentionally want to perform an interaction during a blink. Additionally, \citet{Kirchner2023Eyeball} found eye movements during eyelid closure. Since this might lead to accidental interactions,  we keep the position of the last raycast if the openness of the eye is below a threshold of $0.7$. 
Furthermore, studies suggest a lower blink rate during the use of an HMD \cite{kim2018change}, which may positively influence our system design, leading to fewer accidental selections due to involuntary blinks.
For the pinch gesture, we use the hand tracking system provided by Varjo.
It operates on an Ultraleap~Leap~Motion~Controller~2 and recognizes a gesture once thumb and index finger are held together below the recommended threshold of $0.8$ of Ultraleap\footnote{\url{https://docs.ultraleap.com/xr-and-tabletop/xr/unity/plugin/features/pinch-and-grab-detection.html}}.
To distinguish a selection confirmation from a drag gesture, we check if the hand moved by a certain distance while performing a pinch. Here we found the best minimum distance to be $7$cm and the minimum pinch duration to be $300$ms.\\

To ensure that there are distinctive multimodal inputs for each interaction, we constructed a state machine (cf. ~\cref{fig: states}) such that if the openness of both eyes is below a certain threshold, it will result in a selection confirmation (selection). In contrast, if only a single eye is closed, it will start a drag-and-drop interaction until both eyes are open or closed. To capture the eye positions, we used the highest provided sampling rate of 200 Hz supported by the Varjo XR-4. We decoupled the capture of the gaze positions from the frame rate to avoid missed blinks. Additionally, we provided auditory feedback whenever a participant interacted with a UI element.

\subsubsection{Procedure \& Participants:}
\label{sec: Procedure and Participants}
We recruited participants through our university's study platform, which is mostly used by human-computer interaction and psychology students,
who were compensated with credits. Our exclusion criteria: no pacemaker or past cases of epilepsy, following Tobii guidelines\footnote{\url{help.tobii.com/hc/en-us/articles/212372449-Safety-guidelines}}, and no long, artificial finger nails with nail art. All participants had normal vision or wore contact lenses (glasses were prohibited to ensure optimal eye-tracking performance). Instructions during the study were given in German, requiring all participants to have sufficient language skills.
We performed an a priori power analysis, which indicated a minimum sample size of $n=15$ to detect large effects (Cohen's $d=.8$) with $80\%$ power for a paired-samples t-test with an alpha level of $0.05$.\\

First, participants signed a consent form and filled out a pre-study questionnaire on demographics and XR experience. 
Afterwards, they spent at least 150 seconds inside a small playground environment to familiarize themselves with the interactions and the interface. 
The starting condition was alternated between \emph{Gaze+Pinch} and \emph{Gaze+Blink} for each participant. The participants followed a fixed sequence of tasks as specified in Figure~\ref{fig: task}. For each condition, the task sequence was repeated five times. For task 6 (\emph{Enter password}), the password order for each participant was randomized with balanced Latin squares. The five passwords are listed in the Appendix in~\cref{tab: Passwords}.
We calibrated the eye-tracker before each trial and had the option to re-calibrate the tracking on request.
For optimal blink detection, we performed a manual threshold calibration for the left and right eye on each participant. 
After each condition, participants took off their HMD to rest their eyes. Furthermore, they filled out various questionnaires on a tablet: Simulator Sickness Questionnaire (SSQ)~\citep{kennedy1993simulator}, NASA Task Load Index (NASA Raw-TLX) \citep{hart2006nasa}, System Usability Scale (SUS)~\citep{brooke1996sus}, and short UX Questionnaire (UEQ-S) \citep{schrepp2015user}. After finishing all conditions, participants were asked to fill out a post-study questionnaire with two open questions (cf. Appendix~\cref{tab: Questions Study 1}).
The study took approximately 60 minutes in total.

\subsection{Results}
For the following statistical analysis, except for SSQ data, we tested the normality of our collected data by performing Shapiro-Wilk tests on the differences between our paired measurements for the two interaction techniques: \emph{Gaze+Pinch} and \emph{Gaze+Blink}.
In case the normality assumption was violated, we applied appropriate transformations on the data to achieve a normal distribution and then performed paired samples t-tests. 
We recorded SSQ data at three points of time -- first, a pre-study baseline and once after each condition was completed. We tested for normality of residuals with the Shapiro-Wilk test. Appropriate data transformations were used if the normality assumption was violated. We performed one-way repeated measures ANOVAs and confirmed sphericity using Mauchly's test. If normality of the data could not be assumed, we used the Friedman test.

\subsubsection{Demographics and Prior HMD Experience:}
In total, we collected data from $n=16$ participants (13 women, 3 men), aged from 19 to 39 years (M=$24.9$, SD=$5.05$). Only one is left-handed, while the rest are right-handed. 9 participants have normal and 7 corrected vision. Generally, participants were infrequent users of HMDs with reported usage of: $n=1$ once a month, $n=4$ once a quarter, $n=2$ once every six months, $n=1$ once a year, and $n=8$ less than once a year (including $n=2$ who have never used an HMD before).

\subsubsection{Task Performance Measurements:}
\label{sec: Task Performance Measurements}
We looked at the overall trial measurements and also analyzed the collected data for each interaction task (cf. Study Tasks~\ref{par:studytasks}) within a trial to investigate task-specific differences. For \textit{Selection Error Count}, we considered all selection interactions where a selection confirmation was performed and any UI element except the correct one for progressing the task was hit. We did not consider interactions with no UI involvement as error selections. 

\begin{figure}[t]
\centering
    \begin{subfigure}[t]{0.49\columnwidth}
    \includegraphics[width=\linewidth]{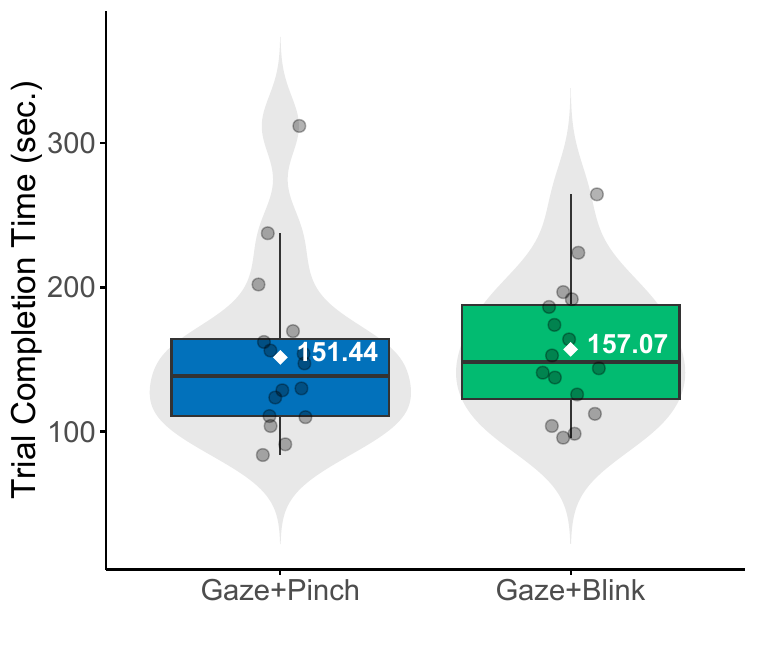} 
    \end{subfigure}
    \begin{subfigure}[t]{0.49\columnwidth}
    \includegraphics[width=\linewidth]{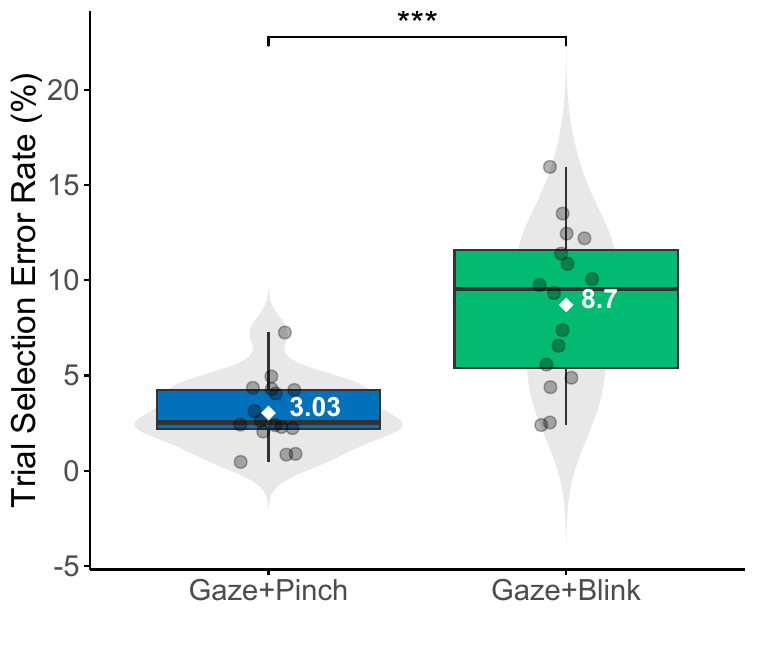} 
    \end{subfigure}
 \caption[]{Overall Trial Measurements, study 1.}
\label{fig:trialdata}

\end{figure}

\paragraph{Overall Trial - Completion Time and Selection Error Rate}
For mean trial completion time, we found no statistically significant difference between \emph{Gaze+Pinch} (M=$151$, SD=$58.8$) and \emph{Gaze+Blink} (M=$157$, SD=$47.5$). 
However, we found a significantly lower mean selection error rate for \textit{Gaze+Pinch} (M=$3.03$, SD=$1.76$) compared to \emph{Gaze+Blink}(M=$8.70$, SD=$4.03$) with t(15) = $-5.23, p< .001$.
As seen in Figure~\ref{fig:trialdata}, the mean selection error rate for \emph{Gaze+Blink} is almost three times as high as for \emph{Gaze+Pinch}, while their mean trial completion time is very similar.

\begin{figure}[t]
\centering
    \begin{subfigure}[b]{0.49\columnwidth}
    \includegraphics[width=\linewidth]{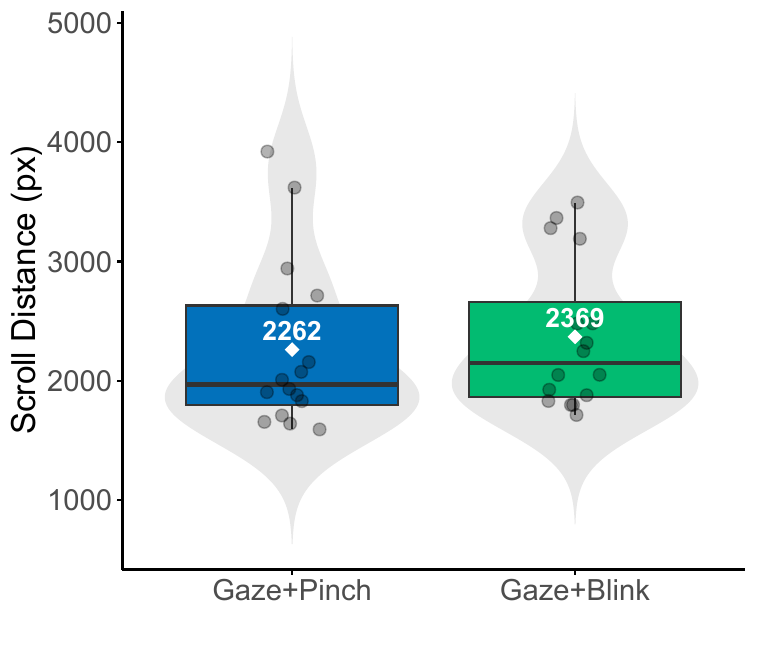}
    \end{subfigure}
    \hfill
    \begin{subfigure}[b]{0.49\columnwidth}
    \includegraphics[width=\linewidth]{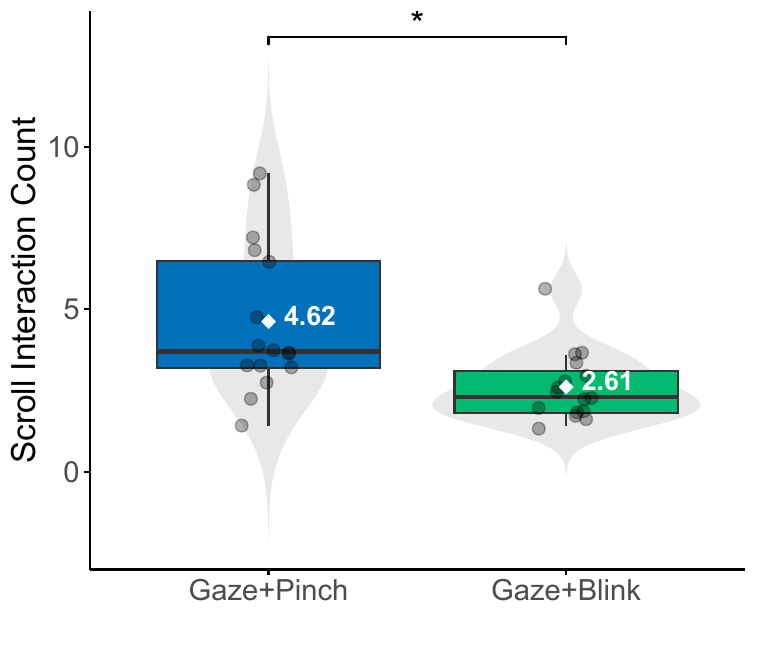}
    \end{subfigure}    
    \vspace{0.5em} 
    \begin{subfigure}[b]{0.49\columnwidth}
    \centering
    \includegraphics[width=\linewidth]{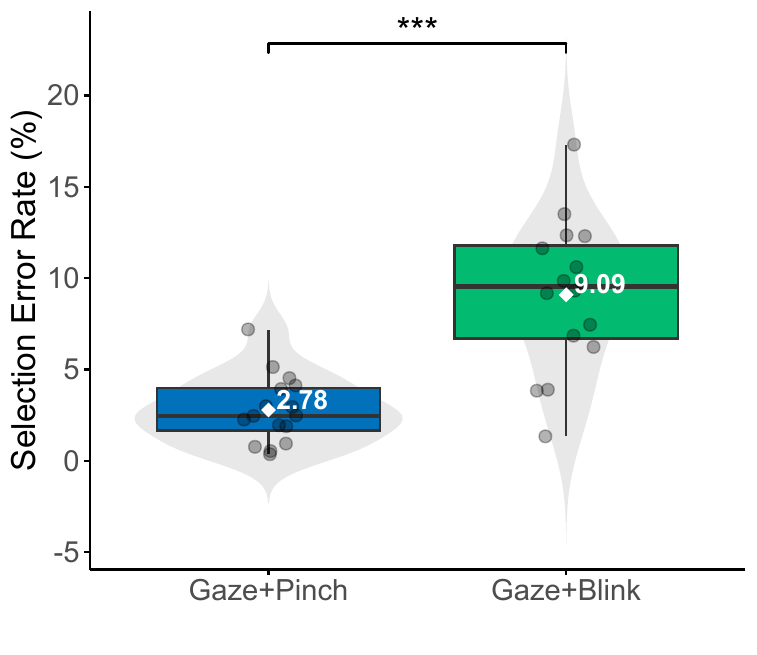}
    \end{subfigure}    
 \caption[]{Settings Menu Task Measurements, study 1.}
\label{fig:settingsdata}
\end{figure}

\paragraph{Settings Menu - Scrolling}
For the settings menu task, while there is no significant difference between mean scrolled distance (in pixels) of \emph{Gaze+Pinch} (M=$2262$, SD=$710$) and \emph{Gaze+Blink} (M=$2369$, SD=$623$), the mean scroll interaction count (each time, a new continuous scroll interaction was started; each new pinch and drag performed on the UI), is significantly higher (t(15) = $2.826, p=.013$) for \emph{Gaze+Pinch} (M=$4.62$, SD=$2.34$) compared to \emph{Gaze+Blink} (M=$2.61$, SD=$1.06$). Although the mean scrolled distance in pixels is very similar, the mean scroll interaction count for \emph{Gaze+Pinch} is almost twice as high as for scrolling with \emph{Gaze+Blink} (cf. Figure~\ref{fig:settingsdata}).
We found a significant difference for mean selection error count for \emph{Gaze+Pinch} (M=$2.78$, SD=$1.85$) and \emph{Gaze+Blink} (M=$9.09$, SD=$4.07$) with t(15) = $-5.72, p< .001$. As seen in Figure~\ref{fig:settingsdata}, \emph{Gaze+Blink} has a mean selection error rate that is three times higher than \emph{Gaze+Pinch}.

\begin{figure}[t]
\centering
    \begin{subfigure}{0.49\columnwidth}
    \includegraphics[width=\linewidth]{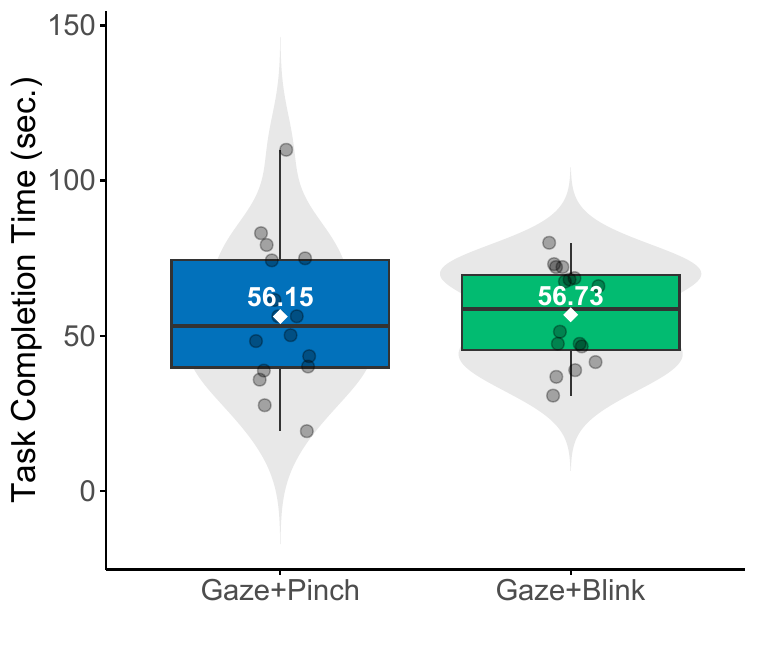} 
    \end{subfigure}
    \begin{subfigure}[b]{0.49\columnwidth}
    \includegraphics[width=\linewidth]{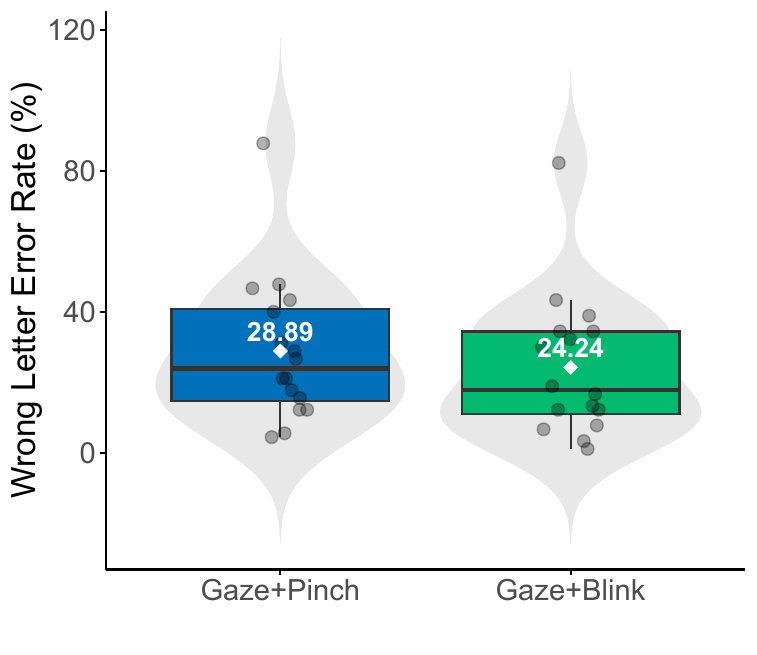} 
    \end{subfigure}
 \caption[]{Wifi Password Keyboard Input Task Measurements, study 1.}
\label{fig:keyboarddata}
\end{figure}

\paragraph{Wifi Password - Text Input}
As illustrated in Figure~\ref{fig:keyboarddata}, the mean keyboard input task completion time is almost the same (\emph{Gaze+Pinch} M=$56.15$, SD=$23.5$; \emph{Gaze+Blink} M=$56.73$, SD=$15.7$) and the mean selection error rate is very close to each other (\emph{Gaze+Pinch} M=$28.89$, SD=$21.0$; \emph{Gaze+Blink} M=$24.24$, SD=$20.5$) in both conditions. The statistical tests revealed no significant differences for text input task completion time and wrong letter selection error rate (amount of selected wrong letters in the given password sequence). 

\begin{figure}[t]
\centering
    \begin{subfigure}[b]{0.49\columnwidth}
    \includegraphics[width=\linewidth]{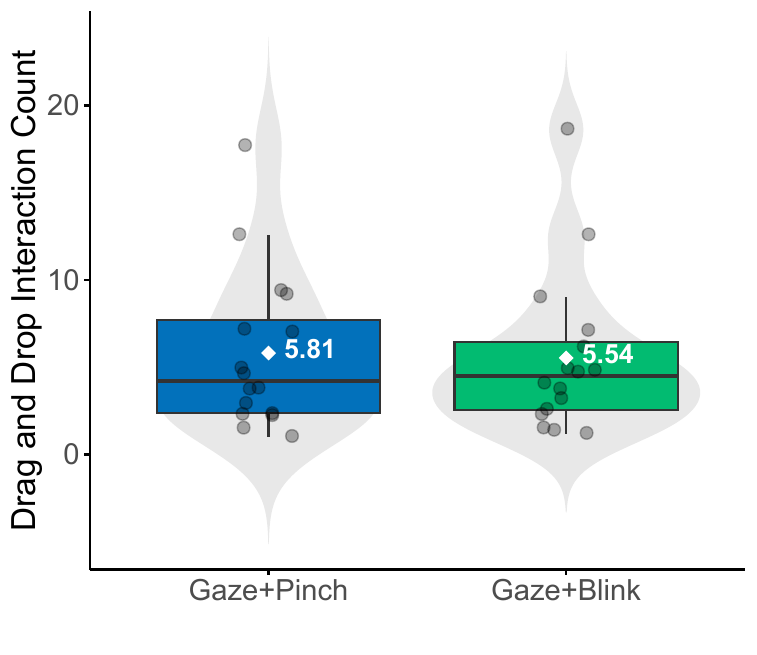}
    \end{subfigure}
    \hfill
    \begin{subfigure}[b]{0.49\columnwidth}
    \includegraphics[width=\linewidth]{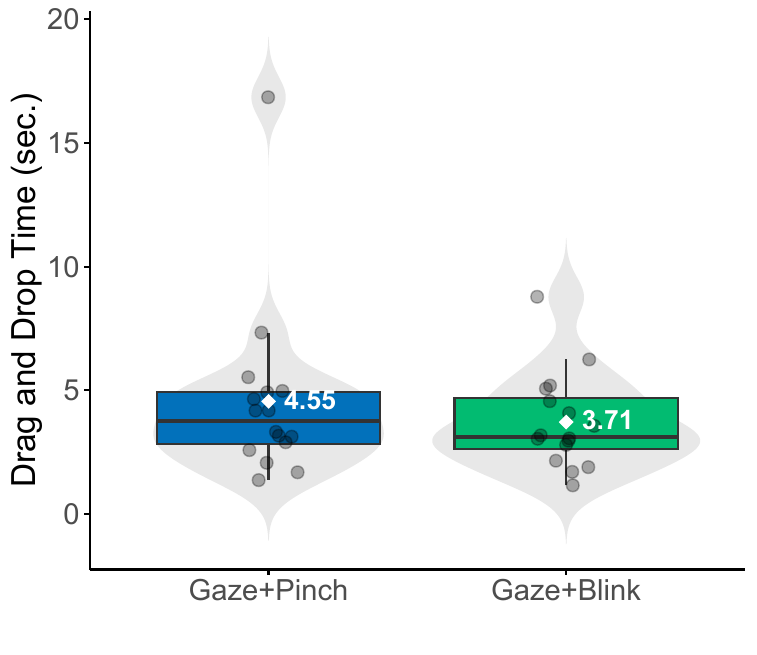}
    \end{subfigure}    
    \vspace{0.5em} 
    \begin{subfigure}[b]{0.55\columnwidth}
    \centering
    \includegraphics[width=\linewidth]{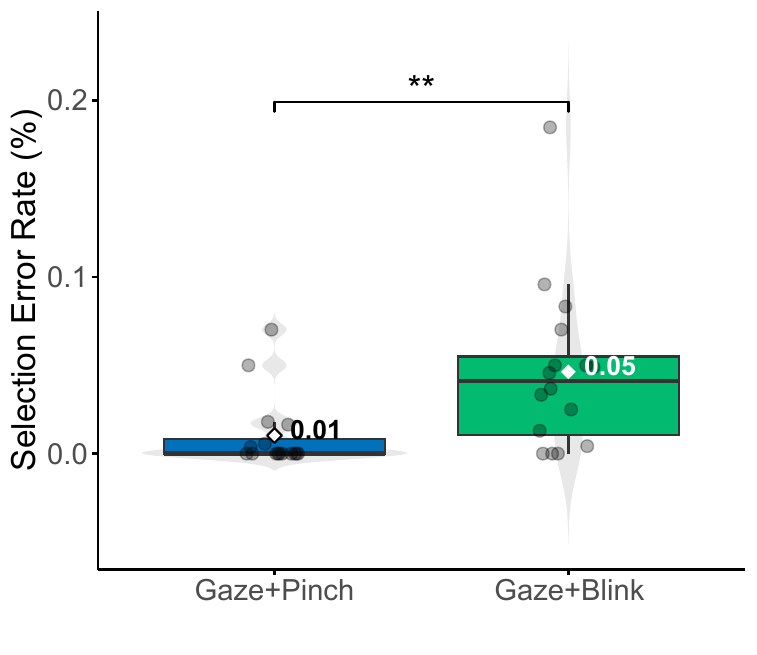}
    \end{subfigure}    
    \caption[]{Messenger Menu drag-and-drop Task Measurements, study 1.}
    \label{fig:messengerdata}
\end{figure}

\paragraph{Messenger - Drag-and-Drop}
We found no significant differences for mean drag-and-drop interaction count (\emph{Gaze+Pinch} M=$5.81$, SD=$4.57$; \emph{Gaze+Blink} M=$5.54$, SD=$4.6$) and mean drag interaction time (\emph{Gaze+Pinch} M=$4.55$, SD=$3.63$; \emph{Gaze+Blink} M=$3.71$, SD=$1.93$). As seen in Figure~\ref{fig:messengerdata}, both measurements are close to each other in terms of means and distribution of the data.
The selection error rate was very low ($<1\%$) for both techniques. Still, \emph{Gaze+Pinch} (M=$0.01$, SD=$0.02$) has a significantly lower mean error rate compared to \emph{Gaze+Blink} (M=$0.05$, SD=$0.05$) with t(15) = $-3.99, p= .0012$.

\subsubsection{Simulator Sickness Questionnaire:}
\label{sec: Simulator Sickness Questionnaire}
\begin{figure}[h]
    \centering
    \includegraphics[width=\linewidth]{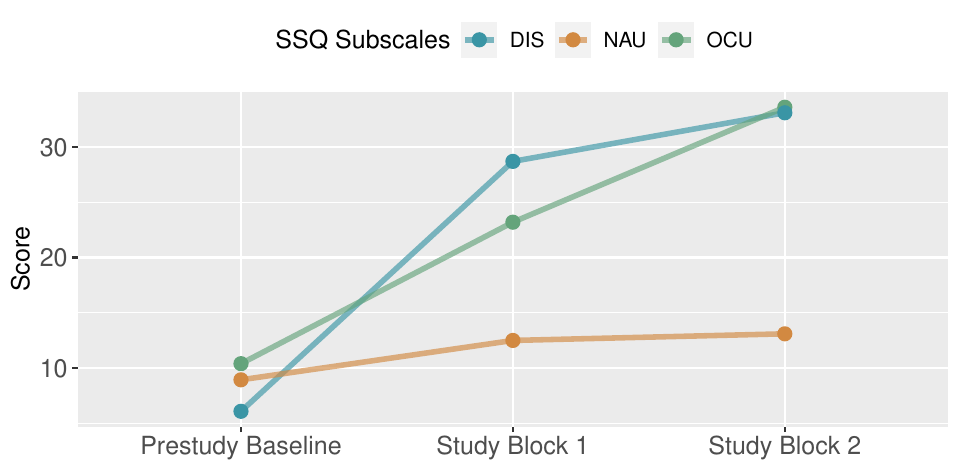} 
 \caption[]{Mean SSQ Subscale Scores, study 1.}
\label{fig:ssq}
\end{figure}

We administered the SSQ before the study (baseline) and after each condition.
The mean SSQ scores are listed in~\cref{tab:ssq-transposed} in the Appendix.
We did not find significant differences for the subscale \textit{Oculomotor Disturbance} and the \textit{Total Simulator Sickness}.
The Friedman tests indicated no significant differences for the subscales \textit{Nausea} and \textit{Oculomotor}, but significant differences for \textit{Disorientation} ($\chi^2(2, N=16) = 16.24, p< .001$). The post-hoc Wilcoxon signed-rank tests, corrected with holm-bonferroni, showed that \textit{Disorientation} was significantly lower for Baseline than for both, Study Block 1 ($p=.011$) and Study Block 2 ($p=.012$). This increase can also be seen in Figure~\ref{fig:ssq}.

\subsubsection{NASA RAW TLX:}
\label{sec: NASA RAW TLX - First Study}
We did not find any significant differences between the two interaction techniques for all subscales and the normalized total workload of the NASA-TLX questionnaire. The mean and SD for each scale are listed for comparison in the Appendix in~\cref{tab:NASATLX-partA-transposed}.

\subsubsection{System Usability Scale:}
\label{sec: System Usability Scale}
The means of the SUS scores for \emph{Gaze+Pinch} (M=$63.4$, SD=$16.8$) and \emph{Gaze+Blink} (M=$63.6$, SD=$14.7$) are very close, and we found no statistical significant differences in SUS scores when comparing both conditions.

\subsubsection{User Experience Questionnaire:}
\label{sec: UEQ}
We used the short version of the UX Questionnaire (UEQ-S) and analyzed the results with the UEQ-S analysis tool. The benchmarks can be found in the Appendix in Figure~\ref{fig: UEQ benchmark graphs}. The \textit{Pragmatic Quality} is rated  bad for both, \emph{Gaze+Pinch} (M=$0.406$, SD=$1.238$) and \emph{Gaze+Blink} (M=$0.375$, SD=$1.158$) in the UEQ-S benchmark. The \textit{Hedonic Quality} is rated good for \emph{Gaze+Pinch} (M=$1.344$, SD=$1.3 90$) and excellent for \emph{Gaze+ Blink} (M=$2.000$, SD=$0.894$). This results in an overall rating of below average for \emph{Gaze+Pinch} (M=$0.875$, SD=$0.838$) and above average for \emph{Gaze+Blink} (M=$1.188$, SD=$0.795$).
We also performed a statistical analysis of the overall rating score, which did not reveal any significant differences between the two techniques. 

\subsubsection{Subjective Feedback:}
\label{sec: Subjective Feedback}
We sorted, grouped, and paraphrased the qualitative user feedback, which we will briefly summarize. Please note that this only provides insights on explicitly stated opinions and does not reflect the complete perspectives and opinions of the participants. Seven participants stated that they preferred \emph{Gaze+Pinch} and nine stated that they preferred \emph{Gaze+Blink}. Generally, participants had contrasting opinions regarding performance and UX for both interaction techniques. While some participants found \emph{Gaze+Pinch} was faster, caused fewer errors, and was easier to use, and they felt that \emph{Gaze+Blink} was more error-prone, other participants were of the opposite opinion. Regarding the UX, some participants reported that \emph{Gaze+Pinch} was more familiar. However, there was also some criticism, including the hand not being tracked, complicated hand-eye coordination, and tense fingers from continuous pinching. \emph{Gaze+Blink} was described as more natural, pleasant and intuitive. Other participants did not agree and instead found it unintuitive, difficult to learn, and more exhausting.

\subsection{Summary}\label{sec: Summary - First Study}
The trial completion time of \emph{Gaze+Blink} does not seem to be negatively impacted by its higher error rate compared to \emph{Gaze+Pinch} (\emph{Hypothesis 1a}). This is also reflected in the performance dimension of the NASA-TLX. The settings menu and drag-and-drop task also show higher error rates for \emph{Gaze+Blink}, while the interaction time does not significantly differ from \emph{Gaze+Pinch}. Therefore, we can only partially confirm our hypothesis. We suspect that involuntary blinks might have caused the higher error rates in \emph{Gaze+Blink}. Paired with insights from the subjective feedback, we speculate that performing multiple consecutive blink interactions may be faster than using pinch gestures, where participants reported hand tracking area limitations and finger strain. These issues could have negatively impacted the trial completion and interaction times for \emph{Gaze+Pinch}. As a result, it was not significantly faster than the more error-prone \emph{Gaze+Blink} in our study. Since the keyboard input times and error rates are comparable, the participants seemed to type equally well using both techniques, \emph{Gaze+Blink} and \emph{Gaze+Pinch}.\\

There were no differences in the perceived workload for both interaction techniques (\emph{Hypothesis 2a}) and in the UX (\emph{Hypothesis 3a}). While the SUS and UEQ-S scores were not significantly different, the UEQ-S benchmarks (cf. Appendix Fig.~\ref{fig: UEQ benchmark graphs}) show that the hedonic quality is rated higher for \emph{Gaze+Blink} than \emph{Gaze+Pinch}. 
The SSQ scores for Nausea and Oculomotor were not significantly higher during the study. Only Disorientation increased significantly while participants used the HMD compared to the pre-study baseline. This could be attributed to the several high ratings for \textit{Fullness of Head}, potentially caused by the fit and relatively high weight of the HMD.\\

In conclusion, \emph{Gaze+Blink} can be a feasible alternative to \emph{Gaze+ Pinch} in terms of speed, overall perceived workload, and UX. Our evaluation revealed that this applies to  both discrete and continuous \emph{Gaze+Blink} interactions. Another notable finding is that the task completion times were not significantly higher for \emph{Gaze+Blink} compared to \emph{Gaze+Pinch}, although the selection error rates were significantly higher. This points in the direction of \emph{Gaze+Blink} compensating for more errors through faster interaction times for each input. 
This motivated our following attempt to reduce the high error rates for \emph{Gaze+Blink} to make it a reliable and more effective interaction technique.
The subjective feedback was also highly varied, with opposing statements regarding perceived errors, speed, and ease of use for both techniques. Different interaction strategies, previous experience with spatial interaction, and individual preferences could have led to differing user experiences, resulting in contrasting feedback. Moreover, possible hand and eye tracking issues could have impacted the UX negatively.\\

Interestingly, \emph{Gaze+Pinch} has a higher scroll interaction count, i.e., the total number of scroll interactions during task \emph{3 -- Scroll to Wi-Fi settings} than \emph{Gaze+Blink} (see Fig. \ref{fig: task}). This suggests that for \emph{Gaze+Pinch}, participants perform several smaller pinch scroll gestures instead of larger hand or arm movements. Fewer head rotations are used for \emph{Gaze+Blink}, but both interaction techniques have comparable scroll distances. This indicates that rotational movement mapping for scrolling with \emph{Gaze+Blink} does not lead to movement mapping issues or increase interaction difficulty, and achieves comparable performance as \emph{Gaze+Pinch}.


\section{Study 2 -- Blink Prediction}
\label{sec: Second Study}
As reported in study 1 (cf.~\cref{sec: Gaze+Blink Interaction}), participants performed more accidental interactions when using \emph{Gaze+Blink}. As stated in \cref{sec: Blink Classification}, involuntary blinks occur quite frequently ($\sim$ 17 blinks per minute) and, therefore, accidental interactions are performed, which can cause frustration to the user.
To mitigate this problem, we propose \emph{Gaze+BlinkPlus} by adapting our technique to mitigate accidental blinks by training a deep-learning model that filters these accidental interactions using the recorded eye-tracking data. We will then use this model to confirm whether a performed blink was voluntary to trigger the interaction accordingly.
While similar work has been done for blink classification \citep{abe2013automatic,sato2015automatic,sato2017automatic}, it has, to the best of our knowledge, not been used for blink-based interaction techniques. Further, as these techniques only use eye images, it might not be possible to utilize these on the device due to privacy reasons, withholding the eye images. Instead, we directly classify blinks using the raw gaze signal along with other eye markers, such as the gaze position or pupil size. The full set of features can be found in~\cref{tab: Features}.
\subsection{Data Collection}
\label{sec: Data Collection}

\subsubsection{Study design:} For our study design, we followed the previous study described in detail in~\cref{sec: First Study -- Study Design}.

\subsubsection{Technical Implication \& Assumptions:} As our \emph{Gaze+BlinkPlus} model requires a trained neural network, we collected additional data for intentional blinks. We designed this as a separate experiment to avoid interference with study 1 on Gaze+Blink due to an additionally required button press that, we assume, will require more cognitive workload. To ensure that all blinks were marked as intentional, we adapted the first experiment (cf.~\ref{sec: First Study -- Study Design}) by requiring participants to confirm if their blink was intentional.
Here, we set out to mark a blink as intentional if the participant simultaneously pressed a button within a margin of 200 ms to fall within the human reaction times \citep{jain2015comparative}. We ensured that participants had to press both the button and simultaneously perform a blink to perform an interaction, to gather reliable ground truth data.

\subsubsection{Participants and Prior HMD Experience:}
Participants were recruited with the same procedure from study 1 (\cref{sec: Procedure and Participants}). We collected data from $n=11$ participants ($4$ women, $7$ men), aged from 22 to 34 years (M=28.3, SD=4.7). $6$ participants have normal and $5$ corrected vision. $n=4$ participants use HMDs more than once a week and $n=3$ once a week, the others are infrequent users with $n=2$ once every quarter and $n=2$ less than once a year. Each participant was required to go through 15 trials (task sequences) with 15 different passwords that can be found in~\cref{tab: All Passwords} in the Appendix.

\subsubsection{Results:}
\label{sec: Results - Data Collection} In total, we collected 6,348,034 data points, including 19,699 blinks that we utilized for model training. During our data collection, we captured 9,221 intentional and 10,508 unintentional blinks. We used 80\% of the participants' data (9 captured sessions) from the data collection as our training set, 10\% as validation set (1 session) and 10\% as test set (1 session), allowing us to validate, train, and test on completely unseen participants. This results in 16,120 training samples (=82\% of total data), 1,581 validation samples (=8\% of total data), and 1,998 test samples (=10\% of total data). On average, we found that participants were performing a blink every 1.6 seconds during our data collection.

\subsection{Methodology}
\subsubsection{Deep-Learning Model:} 
\label{sec: Deep-learning Model}
\begin{figure}[t]
    \centering
    \includegraphics[angle=90,origin=c,width=0.9\columnwidth,trim=2.5cm 0.5cm 2.35cm 1.2cm,clip]{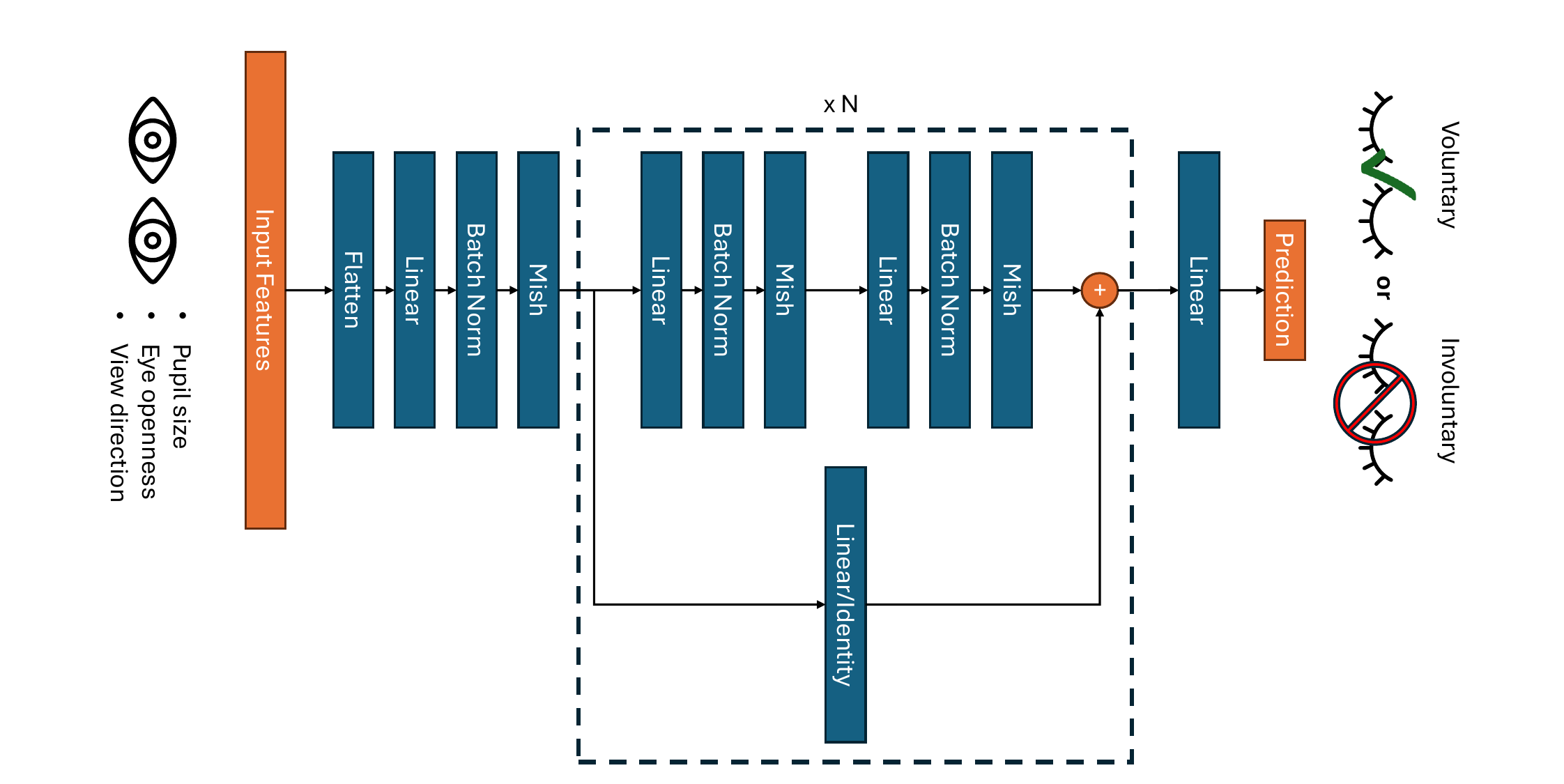}
    \caption{Deep-learning architecture used to predict voluntary/ involuntary blinks. 
    An overview of the model with in- and output dimensions can be found in~\cref{tab: Model} of the appendix.}
    \label{fig: Network}
\end{figure}
While there has been some recent work on time-to-event prediction in the context of VR by \citet{rolff2022saccades,rolff2023deep} that aims to predict when a gaze event will occur and could potentially be adapted to blinks, we do not need this information. Instead, we want to classify at the end of a blink if it was voluntary or involuntary without requiring a dense label at each time step, reducing the computational load required on the system. Our system is designed such that we aim to classify blinks of a wide set of users without the need for extensive retraining or calibration of the model. Hence, we designed our system such that it is abstracted to arbitrary UI elements and input situations, as we do not require specific input of the UI element but rather on general multimodal features of the eye tracker. This results in a multi-class classification problem with two labels: (i) voluntary and (ii) involuntary. Similar to \citet{rolff2023deep}, we utilize historic information provided by the signal for predicting the label and split sequences at event occurrence (end of blink). In our case, we opted for 5000 samples as history that amount to 25 seconds of data. The reasoning behind 25 seconds was the observation that at least two involuntary blinks would be performed when the participant blinks 6 times per minute, which was found to be at the lower end of blinks per minute in VR \citep{kim2022change}. We split the data at the end of a blink. Hence, the last sample of the training sequences is the end of a blink that provides the ground truth (voluntary or involuntary blink). The previous 5000 samples before provide the aforementioned historical context. To avoid overfitting on the data, we randomly shift the end of a blink by 10 samples, making it more robust during inference. Note that this does not influence our input technique during free viewing without any task, as we only require a classification prediction during the interaction with a UI element.\\

As our model, we first use a combination of linear layer, batch normalization \citep{ioffe2015batch} and a Mish \citep{misra2019mish} activation function to transform the high dimensional input data into a lower dimensional space. Afterwards, we use multiple layers of ResNet \citep{he2016deep} blocks, consisting out of two sub blocks with a linear layer, batch normalization layer and a Mish activation function each. To provide optimal gradient flow during training, we add a skip connection around both sub blocks that is either the identify function if the input dimension matches the output dimension or another linear layer otherwise.

To train our model, we use an Adam \citep{kingma2014adam} optimizer with a learning rate of $0.0001$ and train the model for 500 epochs. We also generate a checkpoint after each epoch and another one to store the model with the lowest validation loss on the validation set, which we validate against after each epoch.

\subsubsection{Study Design and Tasks:}\label{par:studytasks2}
We followed the previous study procedure described in~\cref{sec: First Study -- Study Design} with an added third condition based on our deep-learning architecture. We closely followed the study task described in~\cref{par:studytasks}. 
We formulated the following hypotheses to investigate the task performance, perceived workload, and UX between the three conditions \emph{Gaze+Pinch}, \emph{Gaze+Blink}, and \emph{Gaze+BlinkPlus}: 
\begin{enumerate}
    \item[\emph{H1b:}] Task Performance, measured as task completion time and error rate, differs significantly between the three conditions.
    \item[\emph{H2b:}] Perceived workload differs significantly between the three conditions.
    \item[\emph{H3b:}] UX differs significantly between the three conditions.
\end{enumerate} 

\subsubsection{Technical Implication \& Assumptions:}
\label{sec: Technical Implementation and Assumptions - Second Study}
We built our implementation on study 1 (cf.~\cref{sec: Technical Implementation and Assumptions - First Study}), keeping the full experimental setup. As the only extension, we added a third condition where we utilize our model, detailed in~\cref{sec: Deep-learning Model}, to predict voluntary and involuntary blinks. To communicate with the model, we implemented a client/server model that sends the gaze data of the eye-tracker from the Unity application to the server. The Unity application is implemented such that we utilize a background thread, sending the incoming data of each frame to the server over TCP. We capture the data at 200 Hz directly from the device, decoupling the data collection from the frame rate. As data, we send 10 different features listed in~\cref{tab: Features}. Once the end of a blink occurs and 5000 samples (=25 seconds) have been gathered, the server performs a prediction through the model and sends its result back to the application. To listen to the response, we use another background thread, constantly listening for results from the server. Then, if a blink was performed within the last 100ms, we consider that a voluntary blink if the model classified it as such.


\subsubsection{Procedure \& Participants:}
The recruiting followed the same procedure as detailed in~\cref{sec: Procedure and Participants}. However, participants could choose to be compensated with credits or money. 
Our a priori power analysis indicated a minimum sample size of $n=12$ to detect large effects ($\eta_p^2=.4$) with $80\%$ power for an ANOVA (repeated measures, within factors) with an alpha level of $.05$. To match the number of participants of study 1, we again performed our evaluation on $n=17$ participants, amounting to a slightly lower effect size of $\eta_p^2 = 0.33$. We excluded students who participated in the data collection study to avoid contaminating the dataset with known behavior patterns.
%
Study 2 followed the same procedure and used the same questionnaires as the first one, except for the added condition \emph{Gaze+BlinkPlus}. We pseudo randomized the condition order with balanced latin-squares.


\subsection{Results}

\subsubsection{Demographics and Prior HMD Experience:}
We collected data from 17 participants (8 women, 9 men), aged between 20 and 61 years (M=$28.71$, SD=$8.73$). Among them, only $n=2$ participants were left-handed, while the remainder were right-handed. $9$ participants had normal vision and $8$ corrected to normal vision.
Most participants were infrequent users of HMDs with reported usage of: $n=3$ more than once per week, $n=1$ once a month, $n=2$ once every quarter, $n=3$ once every six months, $n=4$ once a year, and $n=4$ less than once a year. 
 
\subsubsection{Task Performance Measurements:}
\paragraph{Overall Trial - Completion Time and Selection Error Rate}
\begin{figure}[t]
\centering
    \begin{subfigure}[t]{0.49\columnwidth}
    \includegraphics[width=\linewidth]{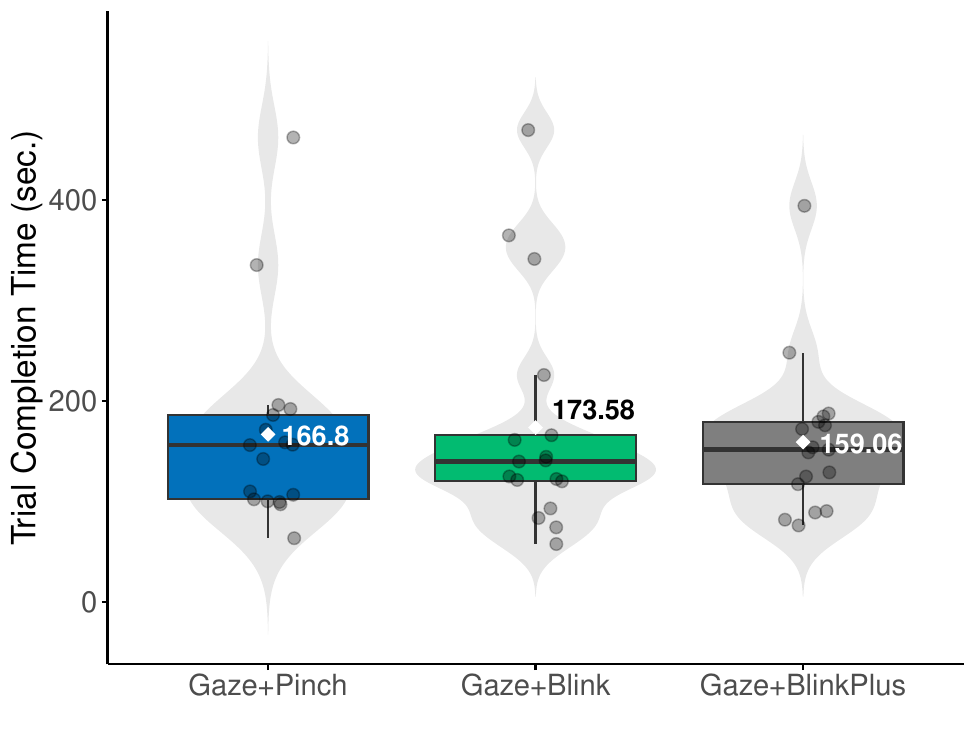} 
    \end{subfigure}
    \begin{subfigure}[t]{0.49\columnwidth}
    \includegraphics[width=\linewidth]{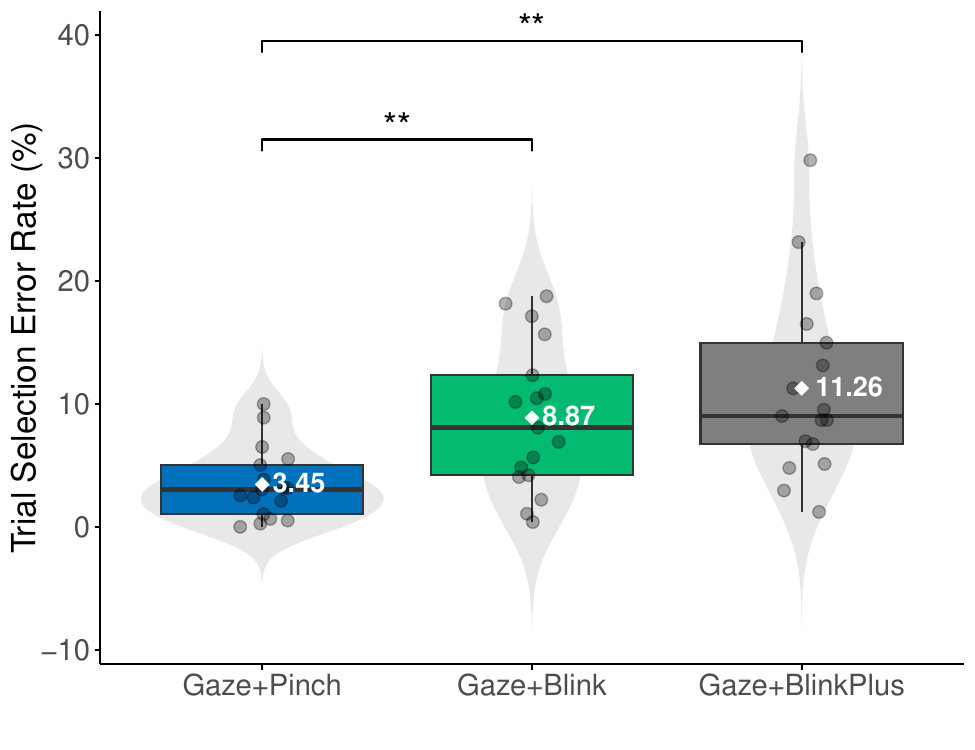} 
    \end{subfigure}
 \caption[]{Overall Trial Measurements, study 2.}
\label{fig:trialdata-2}
\end{figure}
\begin{figure}[t]
\centering
    \begin{subfigure}[t]{0.49\columnwidth}
    \includegraphics[width=\linewidth]{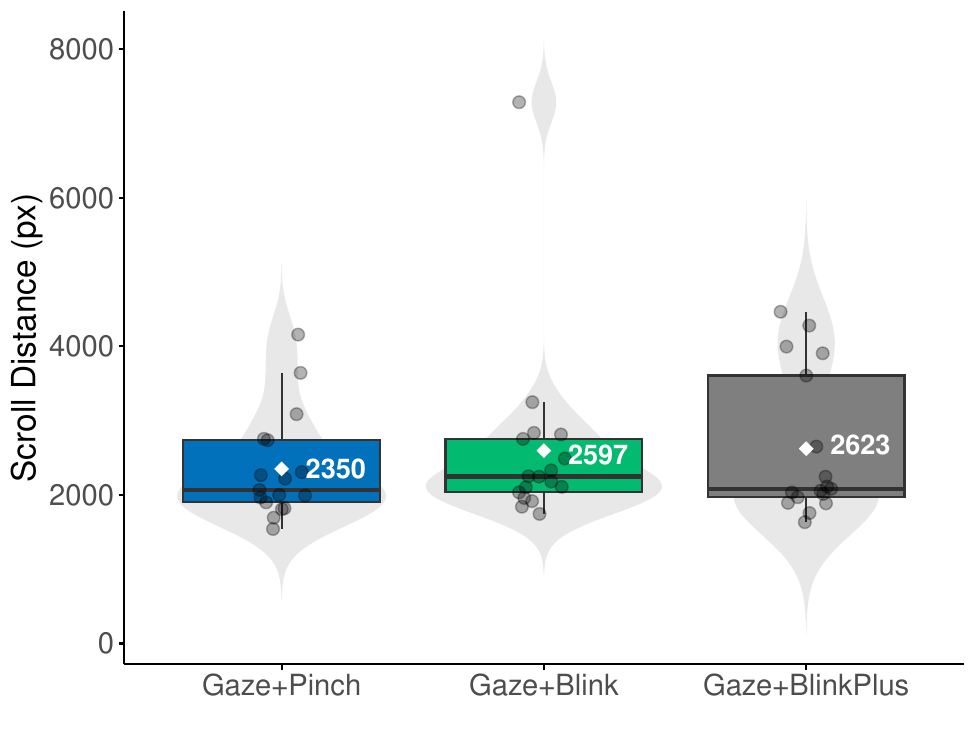} 
    \end{subfigure}
    \begin{subfigure}[t]{0.49\columnwidth}
    \includegraphics[width=\linewidth]{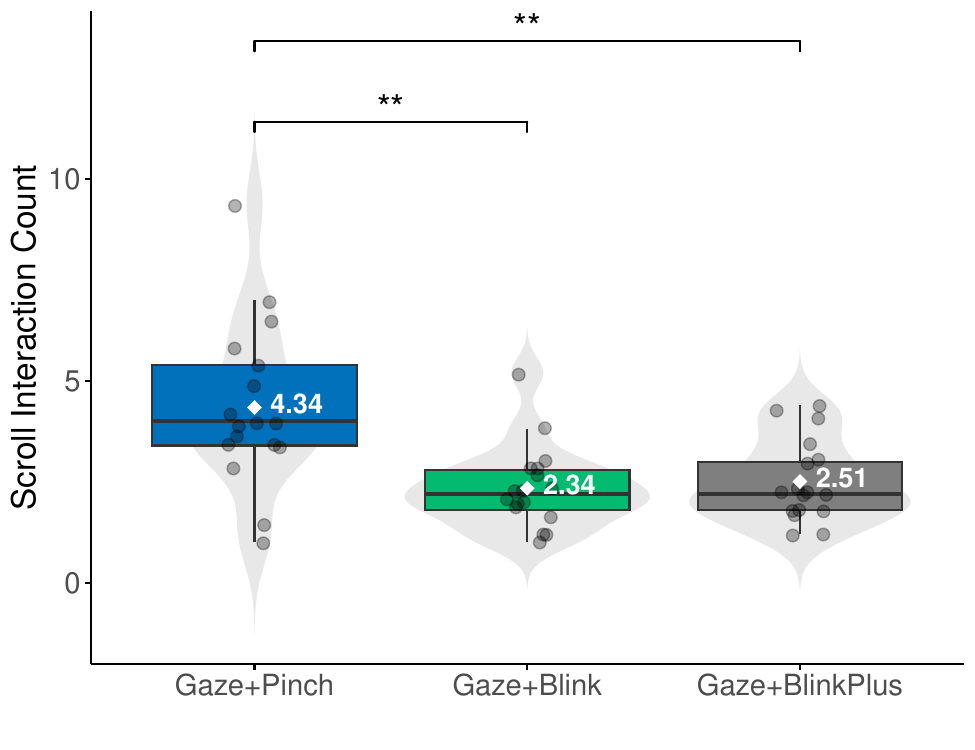} 
    \end{subfigure}
    \vspace{0.5em} 
    \begin{subfigure}[t]{0.49\columnwidth}
    \includegraphics[width=\linewidth]{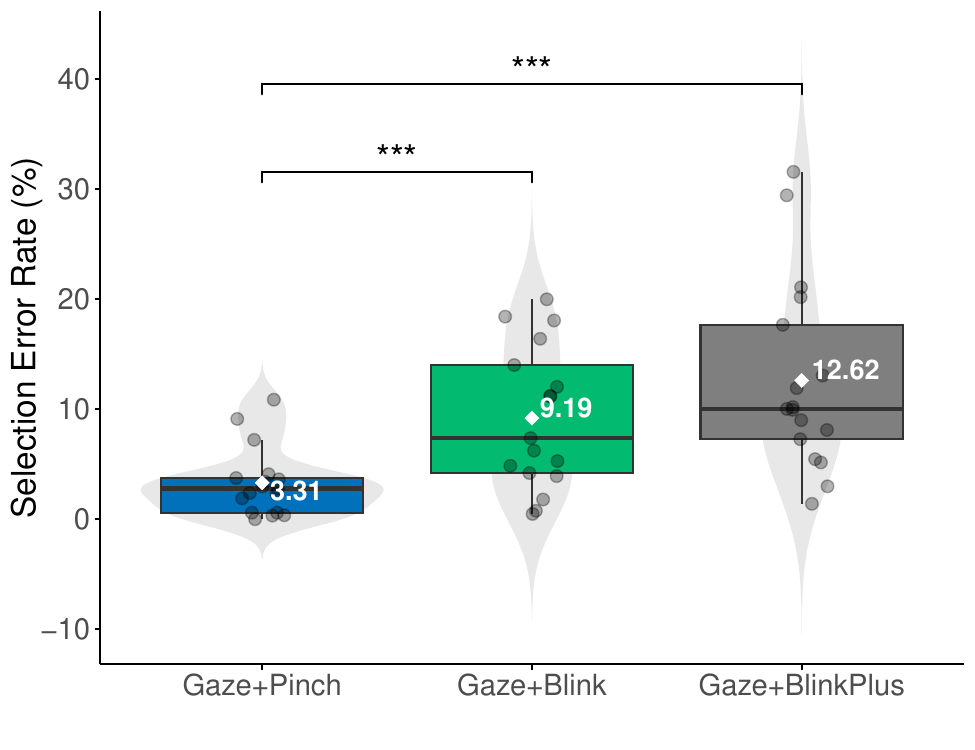} 
    \end{subfigure}
 \caption[]{Settings Menu Task Measurements, study 2.}
\label{fig:settingsdata-2}
\end{figure}

The trial completion time was log-10-transformed to fulfill the normality assumption.
A repeated-measures ANOVA did not yield a significant difference.
For the overall error rate, a repeated-measures ANOVA revealed a significant effect of the selection technique; $F(2,32)=15.20, p<.001, \eta_p^2=.487$.
Pairwise comparisons with Bonferroni-adjusted p-values yielded significant differences between \emph{Gaze+Pinch} and both \emph{Gaze+Blink} ($p=.002$) and \emph{Gaze+BlinkPlus} ($p=.001$), but not between the two blink techniques.
As in study 1, we further examined performance separately for the three tasks.

\paragraph{Settings Menu - Scrolling}
Scroll distance was not significantly different between the interaction techniques according to a Friedman test; $\chi^2(2)=5.77,p=.056$.
Both scroll interaction count and selection error rate for the scrolling task were sqrt-transformed.
Scroll interaction count was additionally corrected using Greenhouse-Geisser because Mauchly's test indicated a violation of the sphericity assumption.
A repeated-measures ANOVA showed a significant effect of the condition on scroll interaction count; $F(1.458,23.334)=12.99, p<.001, \eta_p^2=.448$.
Post-hoc tests with Bonferroni correction revealed significant differences between \emph{Gaze+Pinch} and \emph{Gaze+Blink} ($p=.003$), as well as \emph{Gaze+Pinch} and \emph{Gaze+BlinkPlus} ($p=.005$).
The selection error rate for the scrolling task was also found to be significantly affected by the condition; $F(2,32)=22.33, p<.001, \eta_p^2=.583$.
Again, significant differences were found between \emph{Gaze+Pinch} and \emph{Gaze+Blink} ($p=.002$), as well as between \emph{Gaze+Pinch} and \emph{Gaze+BlinkPlus} ($p<.001$).

\paragraph{Wifi Password - Text Input}
The task completion time for keyboard input and the wrong letter selection error rate were transformed using sqrt and log-10, respectively.
Two repeated-measures ANOVAs did not reveal any significant differences for these measures.

\begin{figure}[t]
\centering
    \begin{subfigure}{0.49\columnwidth}
    \includegraphics[width=\linewidth]{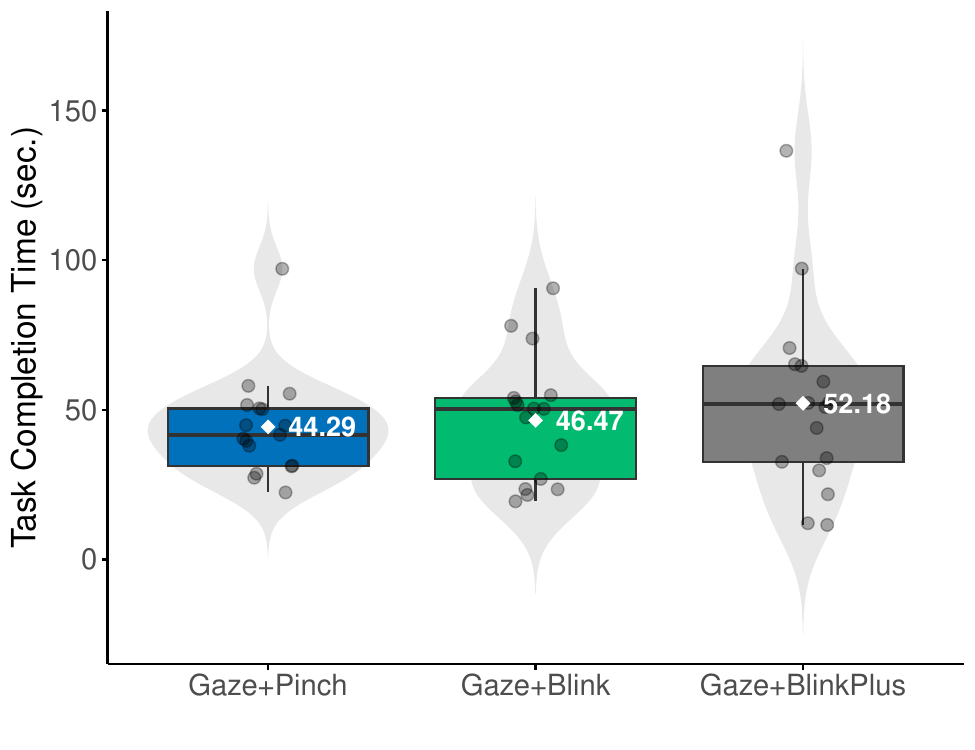} 
    \end{subfigure}
    \begin{subfigure}[b]{0.49\columnwidth}
    \includegraphics[width=\linewidth]{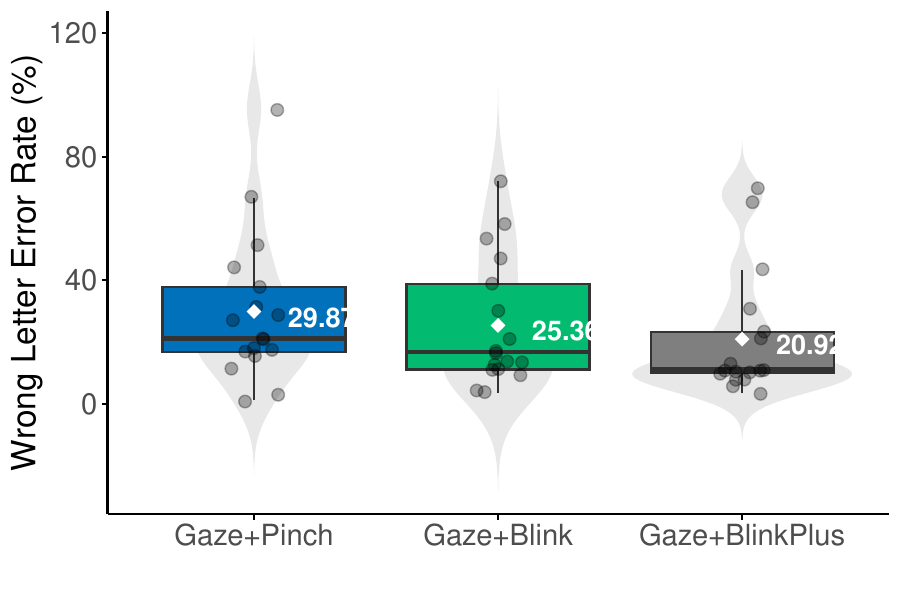} 
    \end{subfigure}
 \caption[]{Wi-Fi Password Keyboard Input Task, study 2.}
\label{fig:keyboarddata-2}
\end{figure}
\begin{figure}[t]
\centering
    \begin{subfigure}[t]{0.49\columnwidth}%
    \includegraphics[width=\linewidth]{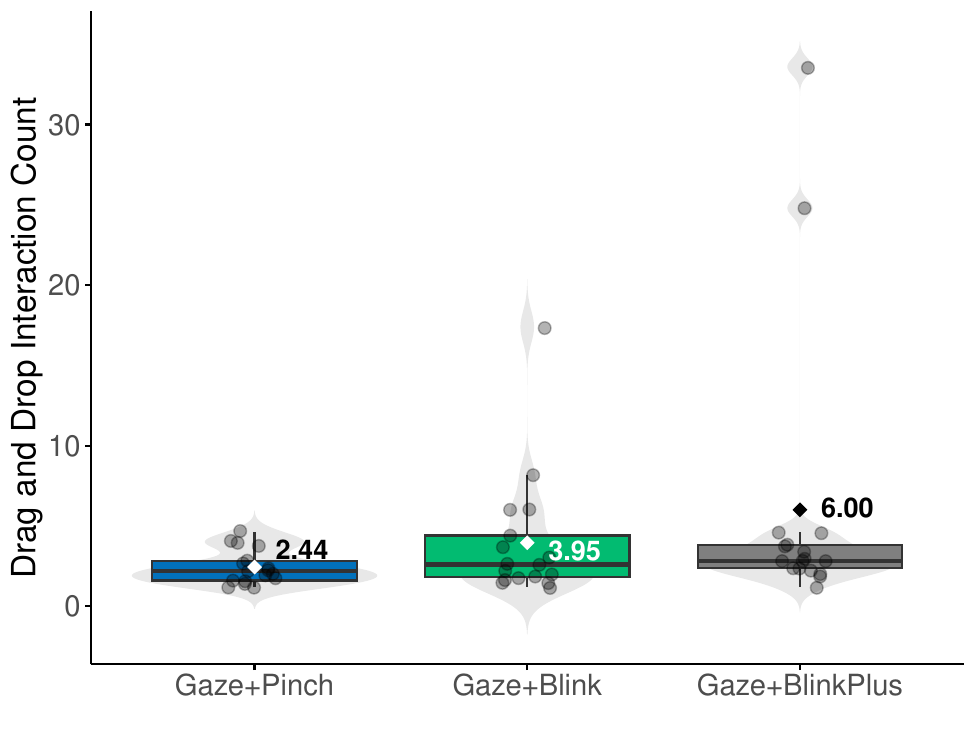}%
    \end{subfigure}
    \begin{subfigure}[t]{0.49\columnwidth}%
    \includegraphics[width=\linewidth]{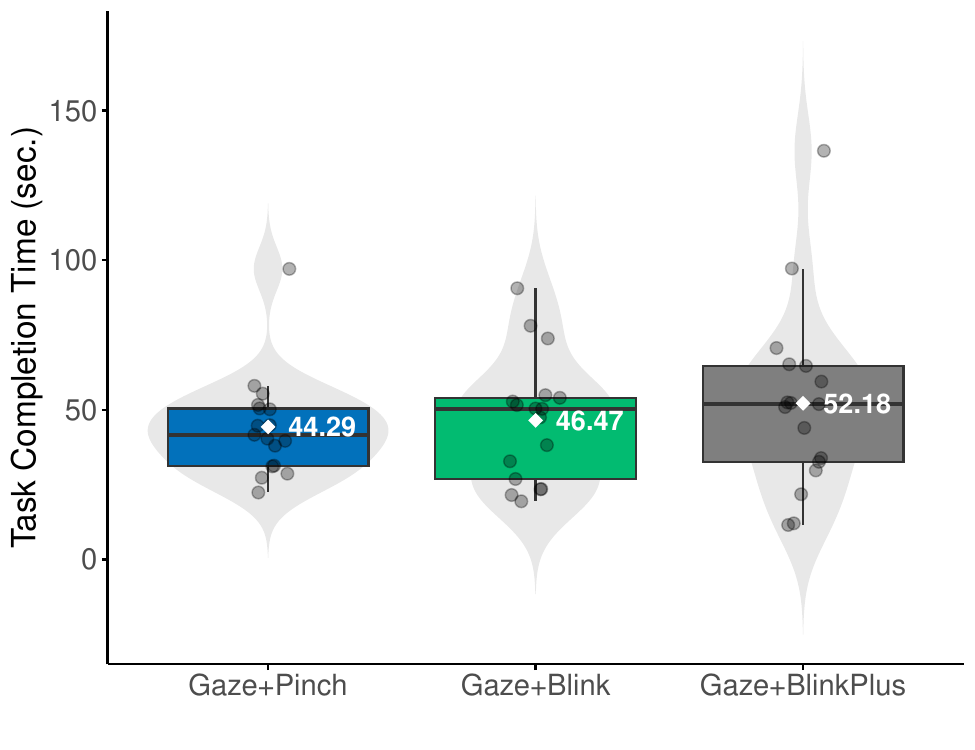}%
    \end{subfigure}
    \vspace{0.5em} 
    \begin{subfigure}[t]{0.49\columnwidth}%
    \includegraphics[width=\linewidth]{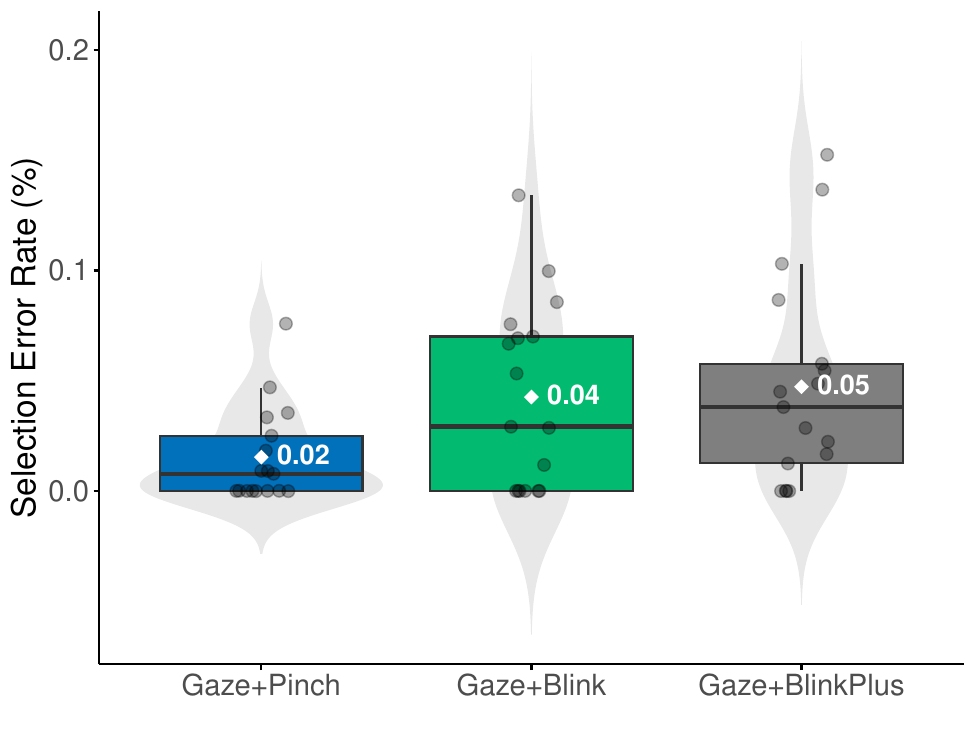}%
    \end{subfigure}%
 \caption[]{Messenger Menu Drag-and-Drop Task Measurements, study 2.}
\label{fig:messengerdata-2}
\end{figure}

\paragraph{Messenger - Drag-and-Drop}
For the drag-and-drop task, no significant differences were found for the performance measures, i.e., selection error rate, interaction count, and interaction time.

\paragraph{Simulator Sickness Questionnaire:}
We followed the same approach as in~\cref{sec: Simulator Sickness Questionnaire} and did not find significant differences on any of the three subscales and for the total SSQ score. Throughout the study, a visual increase in the subscales can be observed, especially for the subscales \emph{Disorientation} and \emph{Oculomotor} (\cref{fig:ssq2}).
The mean scores for each subscale can be found in the Appendix in~\cref{tab:ssq-transposed}. 

\begin{figure}[h]
    \centering
    \includegraphics[width=\linewidth]{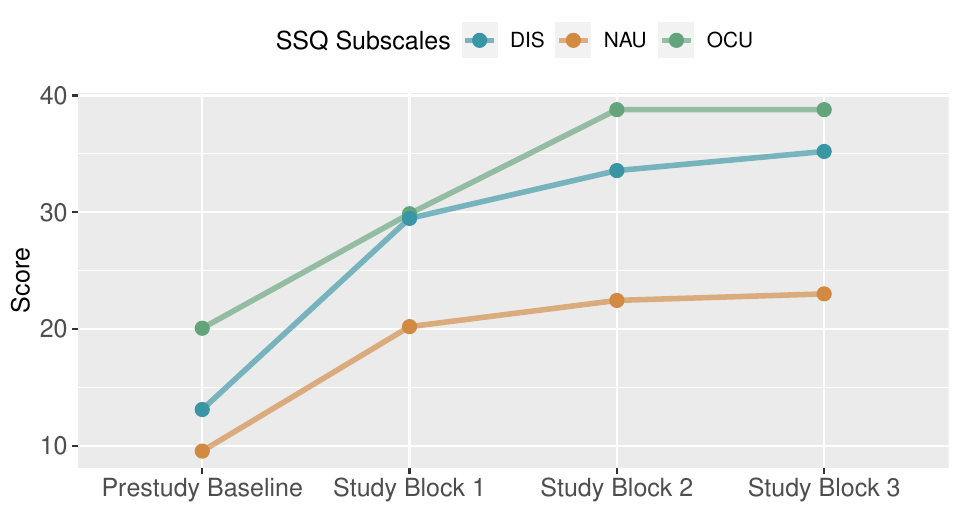} 
 \caption[]{Mean SSQ Subscale Scores, study 2.}
\label{fig:ssq2}
\end{figure}


\subsubsection{NASA RAW TLX:}
Same as in~\cref{sec: NASA RAW TLX - First Study}, we did not find any significant differences between the three interaction techniques for all subscales and the normalized total workload of the NASA-TLX questionnaire. The mean and SD for each scale are listed for comparison in the Appendix in~\cref{tab:NASATLX-partA-transposed}.

\subsubsection{System Usability Scale:}
Same as in~\cref{sec: System Usability Scale}, the SUS scores for \emph{Gaze+Pinch} (M=$61.6$, SD=$16.6$), \emph{Gaze+Blink} (M=$56.9$, SD=$17.8$), and \emph{Gaze+BlinkPlus} (M=$55.4$, SD=$17.0$) are very similar, and we found no statistically significant differences between them.

\subsubsection{User Experience Questionnaire:}
Same as in~\cref{sec: UEQ}, we used the UEQ-S and analyzed the results with the UEQ-S analysis tool. The \textit{Pragmatic Quality} is rated bad for all three conditions, \emph{Gaze+Pinch} (M=$0.30$, SD=$0.994$), \emph{Gaze+Blink} (M=$0.250$, SD=$1.275$), and \emph{Gaze+BlinkPlus} (M=$0.132$, SD=$1.125$). The benchmarks can be found in the Appendix in~\cref{fig: UEQ benchmark graphs}. The \textit{Hedonic Quality} is rated below average for \emph{Gaze+Pinch} (M=$0.779$, SD=$1.208$), good for \emph{Gaze+Blink} (M=$1.309$, SD=$0.990$), and above average for \emph{Gaze+BlinkPlus} ($0.985$, SD=$1.191$). This results in an overall rating of bad for \emph{Gaze+Pinch} (M=$0.544$, SD=$0.725$), below average for \emph{Gaze+Blink} (M=$0.779$, SD=$0.911$), and below average for \emph{Gaze+BlinkPlus} (M=$0.559$, SD=$0.921$).
Again, we performed a statistical analysis of the overall rating score, which did not reveal any significant differences between the three techniques. 


\paragraph{Subjective Feedback:}
\label{sec: Subjective Feedback-2}
Following our previous approach, the participants' statements were sorted and subsequently paraphrased for clarity. 
Seven participants expressed a preference for \emph{Gaze+Pinch}, while eight favored \emph{Gaze+Blink} or \emph{Gaze+BlinkPlus}. Two participants stated that they liked both input techniques equally, with one preferring \emph{Gaze+Pinch} and \emph{Gaze+BlinkPlus}. 
Again, participants had opposing views. \emph{Gaze+Pinch} was described as easier and faster, but some stated it was more error-prone and slow. Some participants expressed that this technique is more pleasant, but others found it annoying, more tedious, and physically demanding. In contrast, the blink interactions were also described as more pleasant and even more intuitive. However, they were also criticized as inconsistent, more tedious, and annoying.

\paragraph{Model benchmark:}
We further evaluated our model on our test set, validating its quantitative performance. As we perform a classification, we utilize four different metrics: Accuracy, Recall, Precision, and F1-Score. Overall, our model can outperform a simple random baseline that guesses the correct class through a random coin throw (cf.~\cref{tab: Benchmark}). We choose this as our baseline because, to the best of our knowledge, there is no state-of-the-art classification algorithm for the task at hand. Further, we could not utilize the thresholding mechanism detailed earlier, as it does not determine if a blink was voluntary or involuntary but rather if a blink happened.

\begin{table}[t]
    \centering
    \caption{Quantitative results on our test set using our model and a simple random guess baseline. As data, we use the samples captured from our data capture study in~\cref{sec: Data Collection}. Better results are highlighted in \textbf{bold}, and a higher value is prefered.}
    \begin{tabular}{r|c|c|c|c}
        \toprule
        Method & Accuracy & Recall & Precision & F1-Score\\ 
        \midrule
        Gaze+BlinkPlus & \textbf{0.76} & \textbf{0.70} & \textbf{0.68} & \textbf{0.67}\\
        \hline
        Baseline     & 0.5  & 0.5 & 0.5 & 0.5\\
        \bottomrule
    \end{tabular}
    \label{tab: Benchmark}
\end{table}

\subsection{Summary}
\label{sec: Summary - Second Study}
%
The results of study 2 confirmed that despite \emph{Gaze+Blink} and \emph{Gaze+BlinkPlus} having a higher error rate per trial, this seems to not negatively impact the trial completion time compared to \emph{Gaze+Pinch} (\emph{Hypothesis~1b}). This is consistent with our findings of study 1.
Same as in study 1, this is also reflected in the performance dimensions of the NASA-TLX that do not significantly differ from each other. 
In accordance to study 1, we found that the settings menu shows higher error rates for both blink-based techniques compared to \emph{Gaze+Pinch}. We could not find any significant differences between \emph{Gaze+Pinch} and our blink-based techniques for the drag-and-drop task (cf.~\cref{fig:messengerdata-2}).
Therefore, we can only partially confirm our hypothesis (\emph{Hypothesis~1b}). With the \emph{Gaze+Blink} and \emph{Gaze+BlinkPlus} techniques, we suspect that participants might have performed a selection task shortly before a scroll task due to the nature of our proposed interaction technique. Here, we believe that participants first focused on an entry before performing the scrolling task, which leads to a rise in unwanted but not involuntary selections; however, we recommend conducting more research on this. Likewise, we found that \emph{Gaze+Blink} and \emph{Gaze+BlinkPlus} improve the scroll interaction count significantly, making them suitable alternatives for scrolling tasks.\\

For the subjective feedback, participants repeated some statements reported by participants of our first experiment. This especially includes the reported tracking area limitations and finger strain. Again, these issues could have negatively impacted the trial completion and interaction times for \emph{Gaze+Pinch}. 
Participants seemed to type equally well using \emph{Gaze+Blink}, \emph{Gaze+BlinkPlus}, and \emph{Gaze+Pinch}, since the keyboard input times and error rates are highly comparable for all interaction techniques. There were no significant differences in perceived workload based on the results of the NASA-TLX questionnaire (\emph{Hypothesis 2b}).
We did not find any significant differences between the three techniques based on the SUS questionnaire and UEQ-S scores. However, the UEQ-S benchmarks revealed that \emph{Gaze+Blink} is rated higher than \emph{Gaze+Pinch} and \emph{Gaze+BlinkPlus} (Appendix~\cref{fig: UEQ benchmark graphs}). These results could not confirm our \emph{Hypothesis 3b}.
Moreover, presenting \emph{Gaze+Blink} and \emph{Gaze+BlinkPlus} in a within-participants design may have indirectly added a bias. Participants may have rated the seemingly identical conditions lower for the hedonic qualities in the UEQ-S. 
Although the SSQ scores increased during the study, we found no significant differences between the SSQ scores in each study block compared to the pre-study baseline.\\


In conclusion, we believe \emph{Gaze+BlinkPlus} and \emph{Gaze+Blink} to be feasible alternatives to \emph{Gaze+Pinch} in terms of speed and overall perceived workload and UX. However, the high error rates during the scroll task warrant a more in-depth analysis. While \emph{Gaze+BlinkPlus} showed the lowest error during the keyboard interaction, we still believe that this can be further improved with additional data collection and a different deep-learning architecture. The subjective feedback was also highly varied with opposing statements. $n=7$ participants preferred one or both of the blink-based techniques (\emph{Gaze+Blink} $n=6$, \emph{Gaze+BlinkPlus} $n=3$) and $n=8$ participants preferred \emph{Gaze+Pinch}. $n=2$ participants stated that they liked both \emph{Gaze+Pinch} and \emph{Gaze+BlinkPlus} equally.
Furthermore, $n=2$ participants stated that they perceived no difference in the \emph{Gaze+Blink} or \emph{Gaze+BlinkPlus} interaction techniques. One participant also would have liked a mixture of blink-based and pinch-based interactions. 
As in study 1, individual differences, preferences, and experiences, as well as tracking issues, might have negatively impacted the UX.


\section{General Discussion}
\label{sec: Discussion}
\paragraph{\textbf{RQ1: How can an interaction model be implemented that covers both discrete and continuous inputs through Gaze+Blink?}} Looking back at our \textbf{RQ1}, we were able to evaluate our designed interaction techniques thoroughly in two user studies and one data collection study. \emph{Gaze+Blink} seems to be robust for performing discrete and continuous interaction tasks. Users had no major issues with intuitively understanding and coordinating the multi-modal input and executing the interaction state transitions. This is also supported in our NASA RAW TLX and SUS results, which did not reveal any significant difference between \emph{Gaze+Blink} and the established selection technique \emph{Gaze+Pinch}.\\

\paragraph{\textbf{RQ2: How does Gaze+Blink compare to Gaze+Pinch for different discrete and continuous UI interactions?}} Reflecting on \textbf{RQ2}, our experimental results indicate that our blink-based techniques achieve comparable performance to the \emph{Gaze+Pinch} technique when tested with realistic interaction tasks in a user interface based on established operating system designs (in our case, reminding in particular of Apple's VisionOS). We found significant differences between some aspects of the evaluated interaction techniques, but they were also comparable in other aspects. Hypothesis 1a and 1b could only be confirmed partially in terms of selection error rates.  While the measured error rates in study 2 were slightly higher for the \emph{Gaze+Blink} and \emph{Gaze+BlinkPlus} conditions, it did not adversely affect the trial completion time compared to \emph{Gaze+Pinch}. 
There were no significant differences in task completion times. Moreover, our results did not confirm hypotheses 2a, 2b and 3a, 3b.
Our findings suggest that \emph{Gaze+Blink} and \emph{Gaze+BlinkPlus} are generally comparable to \emph{Gaze+Pinch} in terms of performance, perceived workload, and UX. We believe that blink-based interaction techniques can be a viable alternative and additional option for facilitating HMD usage for constraint spaces or for users that are unable to perform a pinch gesture. 
This also further supports that our design for \emph{Gaze+Blink} provides a suitable interaction model for discrete and continuous inputs. (\textbf{RQ1}).\\

\paragraph{\textbf{RQ3: What are the advantages and challenges with Gaze+ Blink?}} Based on our findings for \textbf{RQ2}, we can confidently report regarding \textbf{RQ3} that \emph{Gaze+Blink} does provide a viable alternative as a hands-free interaction technique.
One aspect of concern may be the Heisenberg Effect of spatial interaction~\citep{Wolf2020, Wagner2023} that occurs through eye movements when closing one's eyes \citep{kirchner2022real}. While we mitigate this by keeping the gaze steady if the openness falls below a certain threshold, it still may have potentially influenced our results by contributing to a higher error rate.
The contrasting subjective user feedback also revealed challenges for our interaction technique. Some participants criticized that \emph{Gaze+Blink} was more error-prone, inconsistent, and annoying to use. Some especially stated that it was more difficult and unintuitive. Others expressed more positive sentiments on using this technique, expressing the contrary view that \emph{Gaze+Blink} is intuitive, pleasant, and natural in its usage.
The subjective feedback suggests that differences in individual preferences, interaction strategies, and prior experience with XR might influence user perception and experience.\\


\paragraph{\textbf{RQ4: Are there optimization methods to reduce involuntary blinks for Gaze+Blink that could improve this interaction technique?}} We assume that the negative user feedback is caused by involuntary blinks, which leads us to \textbf{RQ4}. Here, we tried reducing involuntary blinks through a deep-learning model, achieving 75\% accuracy on unseen data without user calibration. Yet, this seems still not high enough to reduce the negatively noted aspects.
Additionally, our \emph{Gaze+BlinkPlus} technique may have inadvertently filtered out some voluntary interactions by participants, which, while potentially making the process more intentional, might also have introduced some level of frustration. Yet, we would like to point out that, while not significant, our \emph{Gaze+BlinkPlus} had the smallest error rate during the selection tasks. This suggests that a model capable of supporting users by filtering out involuntary blinks could enhance the overall interaction experience.
However, it proposes an interesting machine learning challenge that should be addressed in future work and will be discussed later in this section.
%
%
%
We will further discuss our paper grouped into 
additional topics: 


\paragraph{\textbf{Deep-learning architecture:}} According to the user feedback, we found the surprising result that supporting involuntary blink suppression through a blink detection mechanism was perceived by some users as beneficial, while others preferred the \emph{Gaze+Blink} method more. We believe that this could be due to differences in predictive performance of our model depending on the participant. This is also somewhat supported when comparing our model benchmark results on the test and validation sets. Here, our model reaches an accuracy of 95\% on the validation set, and an accuracy of 75\% on the test set. As our model was trained on a general population, it can be therefore employed with a 75\% accuracy without requiring extensive retraining or user calibration. However, we also believe that such steps may potentially improve the performance of the model. This indicates that a model calibration phase that fine-tunes the model would be warranted for better predictive performance and can potentially improve the usability and error rate performance of \emph{Gaze+BlinkPlus}. However, we still believe a predictive performance of 75\% on the test set sets a remarkable baseline for future model development. Here, we would like to note again that we solely rely on eye-tracking markers, like pupil size, view direction, or openness, showing the possibility of predicting voluntary and involuntary blinks without direct eye camera access. 

\paragraph{\textbf{Inability to close one eye:}} During our studies, we found that a small percentage of participants have a \emph{unilateral apraxia of eyelid closing} and are therefore unable to close a single eye. This could be due to multiple factors such as genetics \citep{lin2019cerebral}, a stroke \citep{perez2007unilateral}, or a brain lesion \citep{nicoletti2021eyelid} that results in participants being unable to close one eye. It is often that participants are unable to close the eye bilaterally on both eyes, but still can close it unilaterally on a singular eye. Being unable to perform this closing gesture can also be the result of not having learned how to perform the gesture \citep{lin2019cerebral}. Interestingly, the inability to close one eye is often correlated to the dominant eye \citep{laby2011thoughts}, where it is often more challenging for participants to close the dominant eye. In the future, we would like to advise more research on an alternative interaction technique that only utilizes blinks for scroll and drag-and-drop interactions.

\paragraph{\textbf{Blink suppression:}}
Another factor of our system might be blink suppression \citep{bidder1997comparison}. As proposed by \citet{bidder1997comparison} blink suppression is a mechanism similar to saccadic suppression with the same effect on various visual functions, such as reduced ability to perceive a visual stimulus, and an earlier starting suppression of the stimulus 50-100ms before the blink onset \citep{holmqvist2011eye}. Furthermore, research suggests that the full visual acuity is not recovered until 200-500ms after the blink \citep{ehrmann2005novel}, with the amount of suppression depending on amplitude and task \citep{stevenson1986dependence}. This may introduce a potential influence on speed for our system, as users are limited through blink suppression when performing the interaction.

\subsection{Limitations and Future Work}

\paragraph{Inability to close one eye:} As mentioned above (cf.~\cref{sec: Discussion}), it is quite challenging or impossible for some people to close one eye, making it difficult to use our system. This requires further evaluation or adaptation, such that our proposed drag-and-drop gesture can be performed without this requirement. Hence, we would advise more research on alternative solutions for this challenge.

\paragraph{Computational performance \& other data modalities:} Our current solution requires evaluating a neural network at each blink, which, depending on the hardware, might be too time-consuming to run. Therefore, more research is required to find different architectures that would perform well on mobile hardware or directly run on a hardware-based solution. Even though we restricted ourselves to only eye-tracking features without the eye camera due to the privacy inspired design, it would also be interesting to utilize other data modalities for a better predictive performance.

\paragraph{Limited set of interactions:} As the aim of our paper was the introduction and initial evaluation of a novel interaction technique, we did not delve further into 3D object manipulation, including scaling and rotation. 
However, we believe that it is important to further extend our proposed technique for 3D manipulation tasks. As our technique already supports continuos input, we think that there is potential for creating a blink-interaction framework in future work.

\paragraph{Standardized evaluation} We tested our novel technique with common menu UI interaction tasks, which gave us valuable insights on the performance and UX in realistic use case scenarios. As our results revealed that task completion times were not negatively affected by higher error rates for the blink-based techniques, the performance should be further evaluated. Future user studies should investigate interaction times for selection, drag-and-drop and scrolling tasks in a standardized setup to study additional performance metrics, including the throughput.

\paragraph{User calibration:} As briefly mentioned in the discussion, we did not fine-tune the model for each user. This enables us to predict voluntary and involuntary blinks on a wide variety of users, avoiding costly retraining or calibration of the model. However, it might also have had a considerable impact on its predictive performance, increasing the overall error rate and lowering the overall acceptance of the technique. We believe this to be a next step in future work. Another open question is how a threshold calibration for blinks can be integrated reliably. We would further suggest integrating an adaptive threshold algorithm that changes its threshold depending on the visibility of the eyes, as the algorithms for determining the openness may return lower openness values depending on the direction of the eyes (i.e. when looking down).

\paragraph{Accessibility:} Even though we did not particularly design our method for accessibility from the ground up, we think it might be warranted to perform additional accessibility studies. This should be addressed carefully, as it is important to include affected user groups directly in the process. 

\section{Conclusions}

We introduced two innovative interaction techniques, \emph{Gaze+Blink} and \emph{Gaze+BlinkPlus}, which facilitate hands-free spatial interactions in public and constrained environments. Our findings suggest that \emph{Gaze+Blink} interactions are a viable alternative to \emph{Gaze+Pinch} with comparable performance, perceived workload, and UX. \emph{Gaze+Blink} can be implemented on the same type of HMDs that currently support \emph{Gaze+Pinch} without requiring additional tracking hardware. In the future, a combination of these techniques may help to further support privacy, safety, and accessibility for using spatial computing devices, like the AVP.\\

Our comparative analysis indicated that while the overall completion times of \emph{Gaze+Blink} are comparable to those of \emph{Gaze+Pinch}, they exhibit a slightly higher error rate. To address this issue, we developed a novel machine learning-based blink classification system capable of distinguishing between voluntary and involuntary blinks. 
Study 2 revealed 
similar performance on metrics like trial and task completion time, letter selection error rate, or interaction count for \emph{Gaze+Blink}, \emph{Gaze+BlinkPlus}, or \emph{Gaze+Pinch}. While there are still some significant differences in selection error rate in favor of \emph{Gaze+Pinch}, we also found that our techniques excel at other metrics like scroll interaction count. 
Qualitative feedback from participants revealed no definitive preference between the interaction techniques examined. Specifically, 41\% of participants expressed a preference for either \emph{Gaze+Blink} or \emph{Gaze+BlinkPlus}, while 47\% favored \emph{Gaze+Pinch}. Further, 12\% of participants indicated that they valued both input techniques equally. 
Additionally, we introduce a novel approach to accidental blink classification that relies solely on eye-tracking metrics, including eye openness, pupil size, and gaze direction, without necessitating access to eye cameras. This method can serve as a baseline for future advancements in the field of blink interaction techniques. All in all, this leads us to conclude that our proposed mechanisms represent a viable alternative to \emph{Gaze+Pinch}.\\

In summary, our research offers valuable insights into gaze-blink interaction techniques and highlights the challenges inherent in their implementation. We believe that this work lays a solid foundation for further investigations in the realm of blink-based interactions.

\section*{Funding \& Acknowledgements}
The research for this paper was funded by the Deutsche Forschungsgemeinschaft (DFG, German Research Foundation) under Germany's Excellence Strategy – EXC 2176 ‘Understanding Written Artefacts: Material, Interaction and Transmission in Manuscript Cultures’, project no. 390893796. The research was conducted within the scope of the Centre for the Study of Manuscript Cultures (CSMC) at the University of Hamburg.\\

\noindent The research for this paper was funded by the Deutsche Forschungsgemeinschaft (DFG, German Research Foundation) -- project no. 511498220 at the University of Hamburg.\\

\noindent This work has received funding from the European Union's Horizon 2020 research and innovation program under the Marie Sklodowska-Curie grant agreement No 101086206 PLACES.

\section*{Ethics Statement}
All our studies in the paper took place at the Department of Informatics of the University of Hamburg. The ethics committee of the University of Hamburg approved this human-subject research project through an umbrella ethics application that includes this work. We further followed an ethics checklist provided by the committee.

\section*{Data Availability}
The data that support the findings of this study are available from the authors upon reasonable request.

\section*{Disclosure Statement}
No potential conflict of interests was reported by the authors.

\newpage
\section*{Appendix}%
\input{sections/appendix}

\printbibliography


\end{document}

%% file: images/states.tex
\begin{tikzpicture}[node distance=3.5cm, auto]
    \node[state, rectangle, minimum width=1.2cm, initial, fill=uhhblue,text=white] (open) {default};
    \node[state, rectangle, minimum width=1.2cm, fill=uhhblue,text=white] (closed) [below=of open] {selection};
    \node[state, diamond, aspect=3, minimum width=1.2cm, fill=uhhgreen] (dragend) [right=8cm of open] {drag end};
    \node[state, diamond, aspect=3, minimum width=1.2cm, fill=uhhgreen] (drag update) [below right=1.8cm and 3.5cm of dragend] {drag update};
    \node[state, diamond, aspect=3, minimum width=1.2cm, fill=uhhgreen] (dragstart) [below=of dragend] {drag start};

    \path[->] (open) edge node [midway, sloped, above] {one eye closed \& head rotation} (dragstart)
                     edge [bend left] node [midway, right] {both eyes closed} (closed);
    \path[->] (closed) edge [bend left] node [midway, left] {both eyes open} (open)
        (dragstart) edge node [midway, sloped, below] {one eye open} (drag update)
        (dragstart) edge node [midway, left] {both eyes open/closed} (dragend)
        (drag update) edge [loop right] node [midway, right] {one eye open} (drag update)
        (drag update) edge node [midway, sloped, above] {both eyes open/closed} (dragend)
        (dragend) edge node [midway, above] {} (open);
\end{tikzpicture}

%% file: images/studyprocedure.tex
\begin{tikzpicture}
  \tikzset{
    bound/.style={
      draw,
      minimum height=2cm,
      inner sep=1em,
    },
    arrow/.style={
      draw,
      minimum height=1cm,
      inner sep=1em,
      shape=signal,
      signal from=west,
      signal to=east,
      signal pointer angle=110,
    }
  }
  \begin{scope}[start chain=transition going right,node distance=-\pgflinewidth]
    \node[arrow,on chain,align=center, fill=orange!20] {1. Home screen\\\phantom{(selection)}};
    \node[arrow,on chain,align=center, fill=orange!25] {2. Open settings\\(selection)};
    \node[arrow,on chain,align=center, fill=orange!30] {3. Scroll to Wi-Fi\\settings (scroll)};
    \node[arrow,on chain,align=center, fill=orange!35] {4. Open Wi-Fi\\settings (selection)};
    \node[arrow,on chain,align=center, fill=orange!40] {5. Select Wi-Fi\\(selection)};
    \node[arrow,on chain,align=center, fill=orange!45] {6. Enter password\\(selection)};
  \end{scope}

  \begin{scope}[yshift=-2cm,start chain=transition going right,node distance=-\pgflinewidth]
    \node[arrow,on chain,align=center, fill=orange!50] {7. Close Wi-Fi\\settings (selection)};
    \node[arrow,on chain,align=center, fill=orange!55] {8. Home Screen\\\phantom{(selection)}};
    \node[arrow,on chain,align=center, fill=orange!60] {9. Open messenger\\(selection)};
    \node[arrow,on chain,align=center, fill=orange!65] {10. Open message\\(selection)};
    \node[arrow,on chain,align=center, fill=orange!70] {11. Open gallery\\(selection)};
    \node[arrow,on chain,align=center, fill=orange!75] {12. Send image\\(drag \& drop)};
  \end{scope}
\end{tikzpicture}

%% file: sections/appendix.tex
\section{Machine Learning Features}
In \cref{tab: Features}, we list the features used for classification of voluntary and involuntary blinks. Additionally, we specify the exact layer sizes of our model in~\cref{tab: Model} for reproducibility. 
We use an input layer for compressing the input from a high dimensional vector into a lower space using a linear layer and process the data through the ResNet \citep{he2016deep} modules (c.f.~\cref{fig: Network} and \cref{tab: Model}). 

\begin{table}[H]
    \caption{Features that we captured during our data collection study and are used to train our model.} 
    \resizebox{0.95\columnwidth}{!}{
    \input{tables/features}
    }
    \label{tab: Features}
\end{table}

\subsection{Model Details}
\begin{table}[H]
    \caption{A detailed overview of our model shown in Figure~\ref{fig: Network}.}
    \resizebox{0.95\columnwidth}{!}{
    \input{tables/architecture}
    }
    \label{tab: Model}
\end{table}
\FloatBarrier

\section{User Studies}

\subsection{Technical Implementation}
The angular sizes of the UI elements are listed in~\cref{tab:ui-angular}. 
Moreover, we added a list of all passwords used during our first and second studies (c.f.~\cref{tab: Passwords}), and all passwords used for data collection (c.f.~\cref{tab: All Passwords}). The passwords were generated at random using a password generator. We further count the number of shifts required to insert the password and the unit Euclidean distance of each word, where adjacent characters have a distance of 1.

\begin{table}[H]
    \centering
    \caption{Table of angular sizes of the UI elements}
    \resizebox{0.95\columnwidth}{!}{
    \begin{tabular}{l|c|c}
        \toprule
         Selectable UI Elements & Width & Height \\
         \midrule
         Main UI Elements & & \\
         \hspace{1em}Main Menu Button & $3.95^\circ$ & $3.95^\circ$\\
         \hspace{1em}Close Window Button & $2.05^\circ$ & $2.05^\circ$\\
         Settings Menu & &\\
         \hspace{1em}Settings List Elements (WLAN, ...) & $13.14^\circ$ &  $1.96^\circ$\\
         \hspace{1em}Wi-Fi Network Menu Bar & $20.25^\circ$ & $2.02^\circ$\\
         \hspace{1em}Wi-Fi Toggle Button & $3.38^\circ$ & $1.56^\circ$\\
         \hspace{1em}Password Input Field & $18.06^\circ$ & $2.02^\circ$\\
         \hspace{2em}Cancel / Confirm Button & $3.93^\circ$ & $1.06^\circ$\\
         Keyboard & & \\
         \hspace{1em}Keyboard Letter Key & $2.71^\circ$ & $2.71^\circ$\\
         \hspace{1em}Keyboard Shift Key & $3.79^\circ$ & $2.71^\circ$\\
         \hspace{1em}Keyboard Space Key & $16.43^\circ$ & $2.71^\circ$\\
         Message Menu & &\\
         \hspace{1em}Message Contact List Element & $14.91^\circ$ & $3.07^\circ$\\
         \hspace{1em}Send Image Button & $1.66^\circ$ & $1.66^\circ$\\
         \hspace{1em}Image Thumbnail & $5.52^\circ$ & $5.52^\circ$\\
         \hspace{1em}Drag-and-Drop Input Field & $22.23^\circ$ & $2.33^\circ$\\
         \bottomrule
    \end{tabular}
    }
    \label{tab:ui-angular}
\end{table}

\begin{table}[H]
    \centering
    \caption{All passwords used during our data collection study (c.f.~\cref{sec: Data Collection}).}
    \resizebox{0.95\columnwidth}{!}{
    \input{tables/allpassword}
    }
    \label{tab: All Passwords}
\end{table}

\begin{table}[H]
    \centering
    \caption{All passwords used during our first and second study (c.f.~\cref{sec: First Study} and~\cref{sec: Second Study}).}
    \input{tables/passwords}
    \label{tab: Passwords}
\end{table}

\subsection{Questionnaires}
In addition to standardized questionnaires (SSQ, NASA TLX, and UEQ-S) we used two open questions during our studies that are listed in~\cref{tab: Questions Study 1}.

\begin{table}[H]
    \centering
    \caption[Open questions used in our questionnaires. We used DeepL for translation.]{Open questions used in our questionnaires. We used DeepL\footnotemark{} for translation.}
    \resizebox{\columnwidth}{!}{
    \begin{tabular}{p{0.05\columnwidth}|p{0.53\columnwidth}|p{0.53\columnwidth}}
       \toprule
       Q & Native Question & Translated Question \\
       \hline
       Q1 & ``Welche der beiden Eingabemethoden hat Dir mehr gefallen? Bitte begründe Deine Antwort kurz.'' & Which of the two input methods did you like more? \\
       \hline
       Q2 & ``Was ist Dir bei den Eingabemethoden besonders aufgefallen?'' & What did you particularly notice about the input methods? \\
       \bottomrule
    \end{tabular}
    }
    \label{tab: Questions Study 1}
\end{table}
\footnotetext{\url{deepl.com/en/translator}}

\subsection{User Study Results}
\begin{figure}[H]
    \centering
    \begin{subfigure}[t]{0.47\columnwidth}
        \centering
        \includegraphics[width=\textwidth]{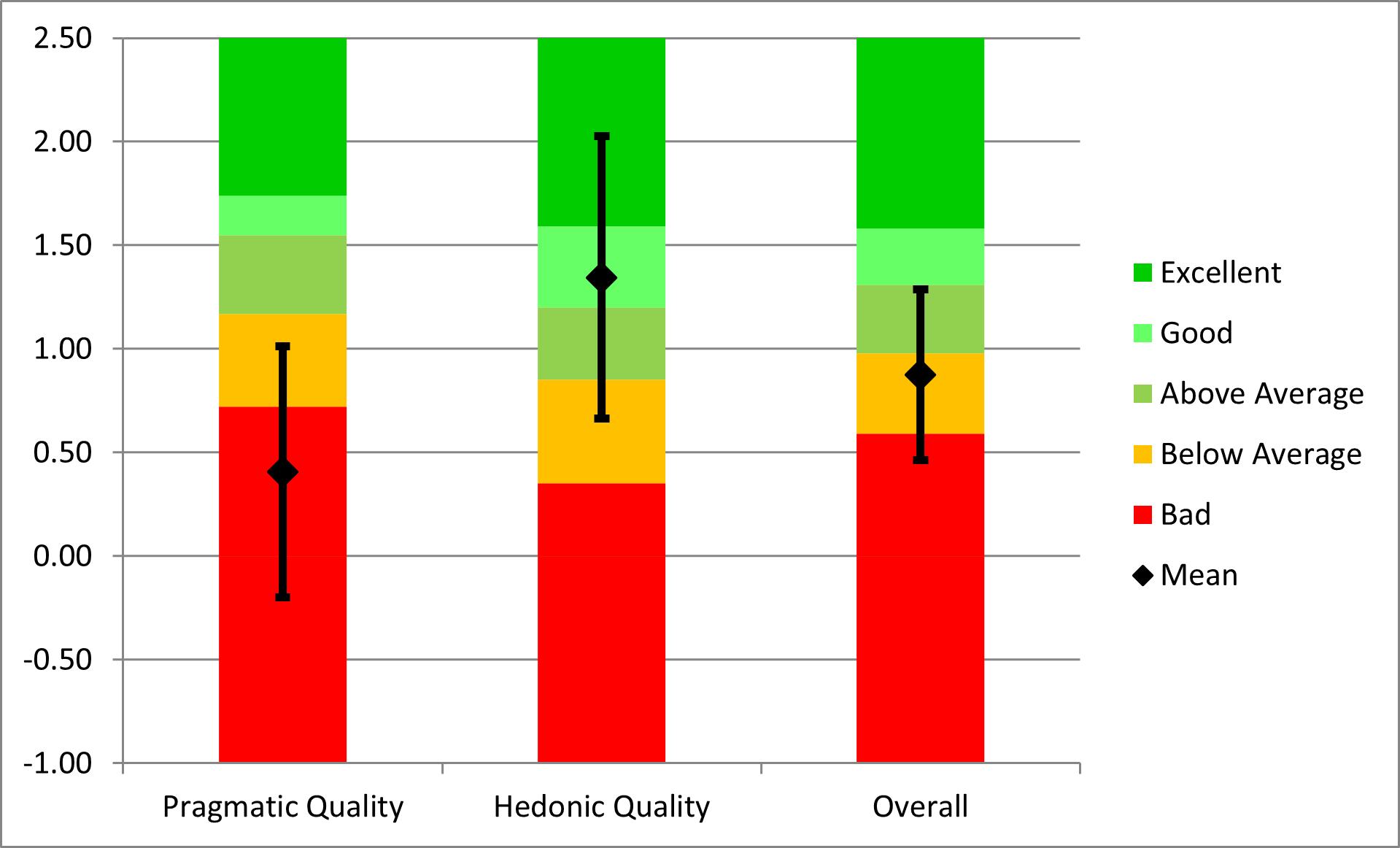}
        \caption{
            UEQ-S benchmark for \textit{Gaze+Pinch} measured during study 1.
        }
    \end{subfigure}%
    \hspace{1em}
    \begin{subfigure}[t]{0.47\columnwidth}
        \centering
        \includegraphics[width=\textwidth]{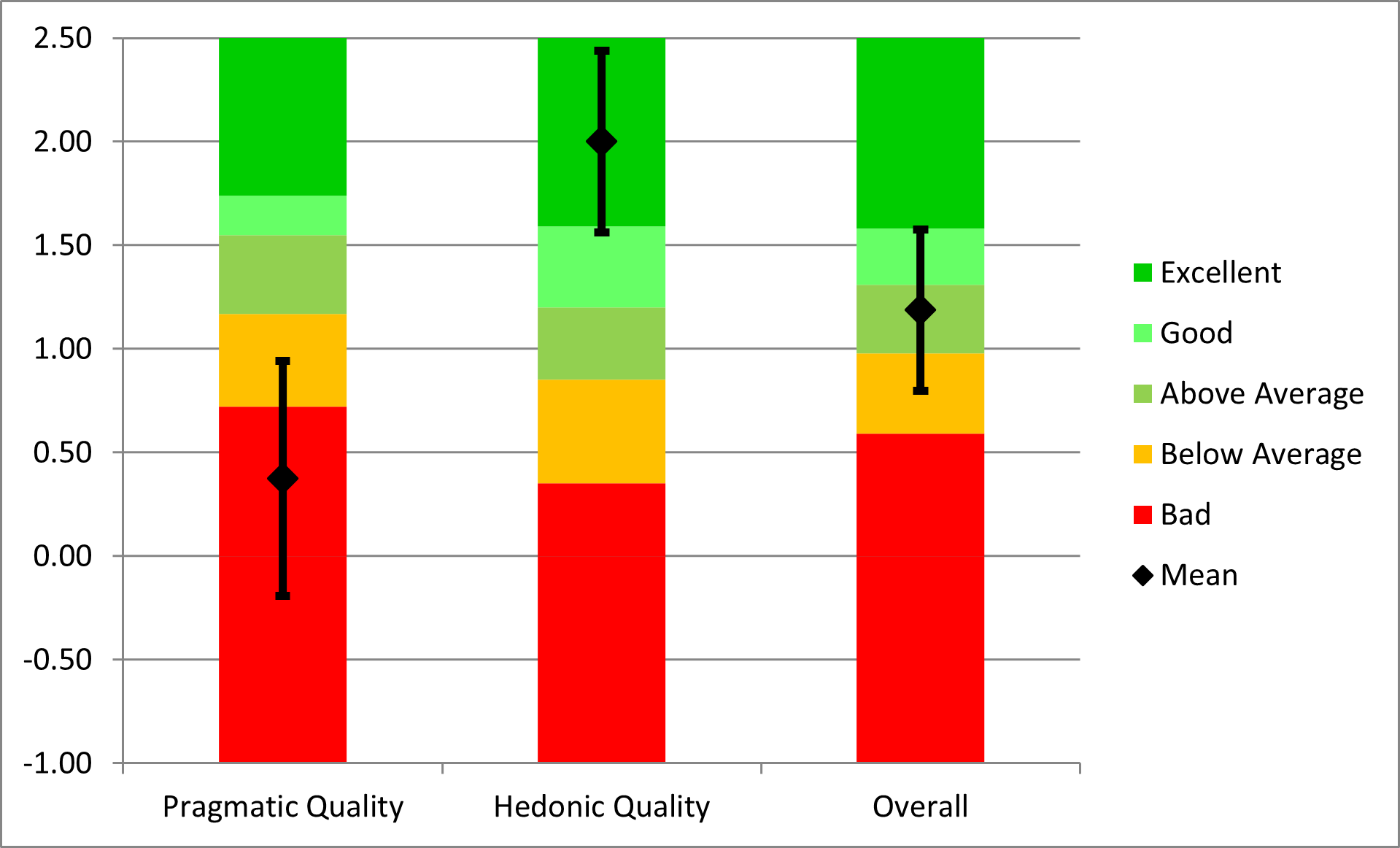}
        \caption{
            UEQ-S benchmark for \textit{Gaze+Blink} measured during study 1.
        }
    \end{subfigure}
    
    \begin{subfigure}[t]{0.47\columnwidth}
        \centering
        \includegraphics[width=\textwidth]{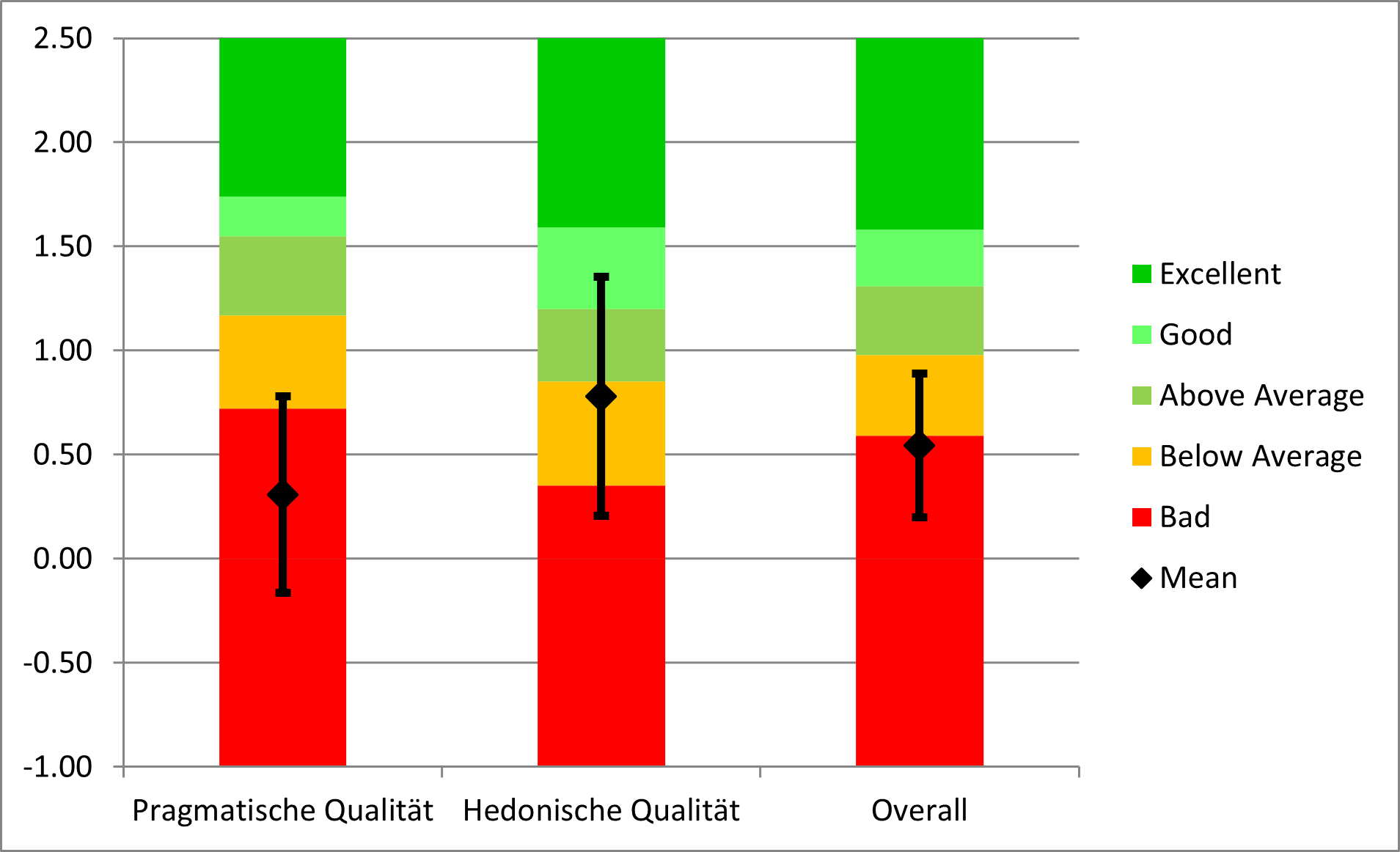}
        \caption{
            UEQ-S benchmark for \textit{Gaze+Pinch} measured during study 2.
        }
    \end{subfigure}%
    \hspace{1em}
    \begin{subfigure}[t]{0.47\columnwidth}
        \centering
        \includegraphics[width=\textwidth]{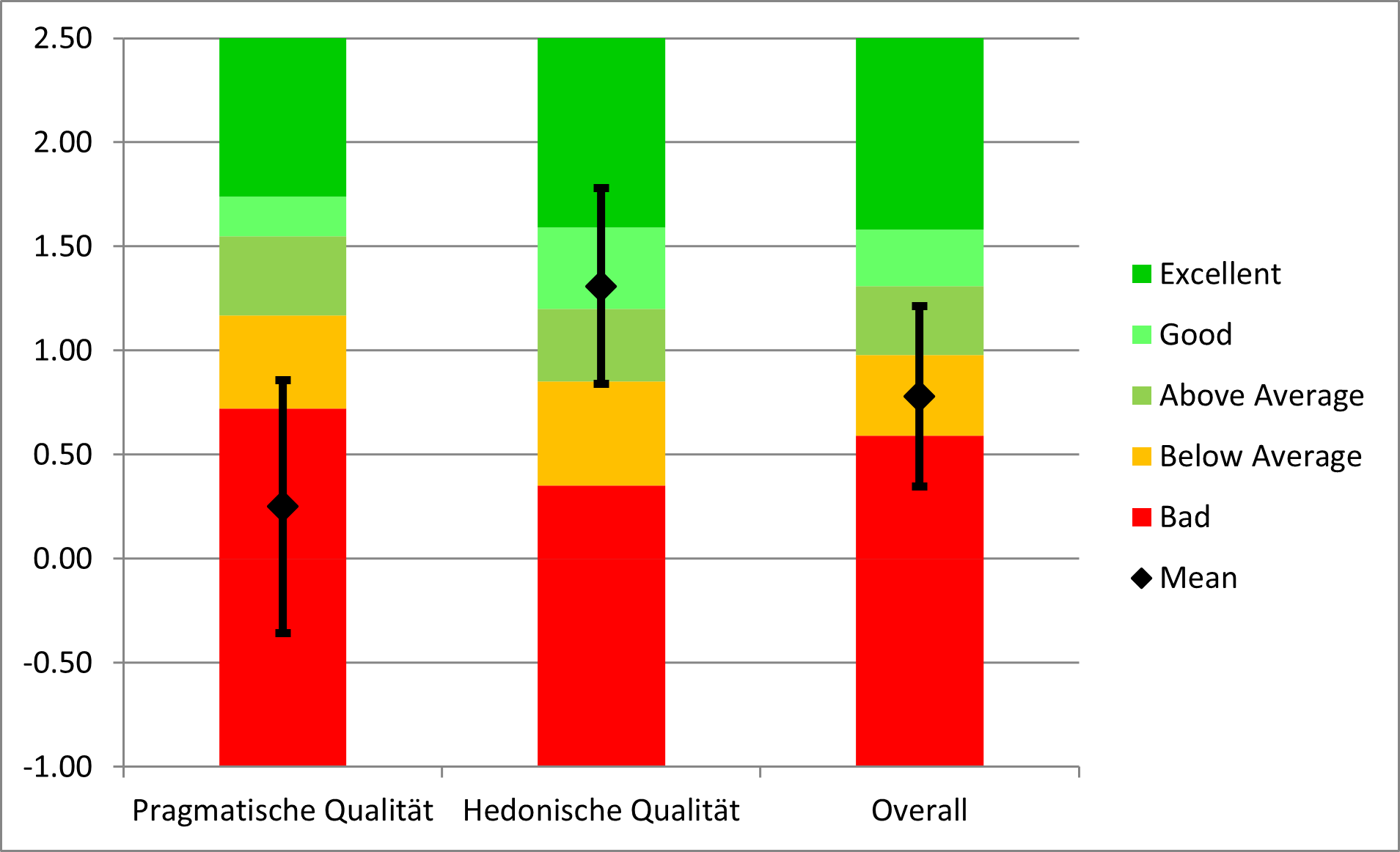}
        \caption{
            UEQ-S benchmark for \textit{Gaze+Blink} measured during study 2.
        }
    \end{subfigure}%
    \hspace{1em}
    \begin{subfigure}[t]{0.47\columnwidth}
        \centering
        \includegraphics[width=\textwidth]{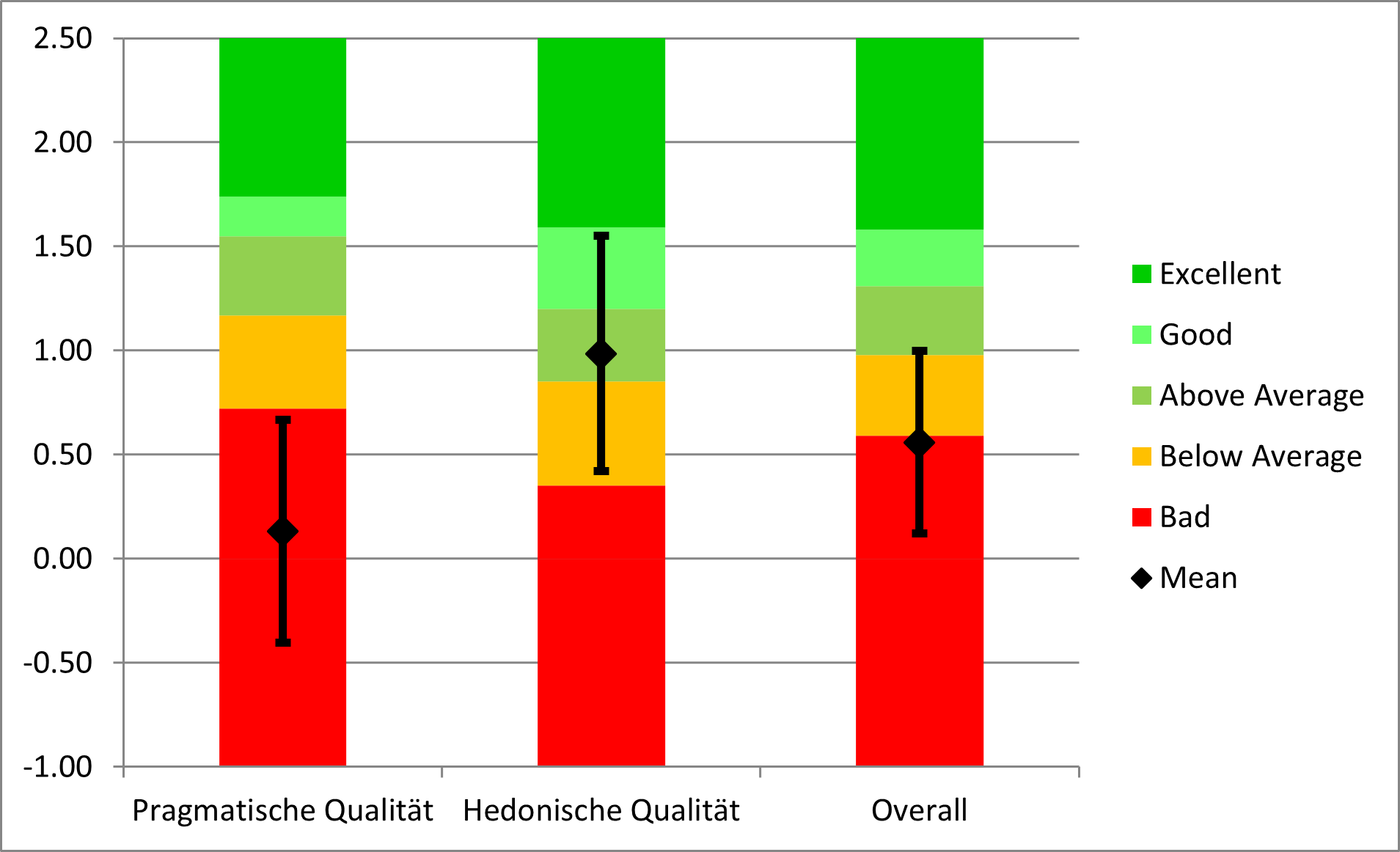}
        \caption{
            UEQ-S benchmark for \textit{Gaze+BlinkPlus} measured during study 2.
        }
    \end{subfigure}
    \caption{UEQ-S benchmark graphs}
    \label{fig: UEQ benchmark graphs}
\end{figure}

We added the SSQ scores for our first and second user study as tables in~\cref{tab:ssq-transposed}. We also listed the NASA RAW TLX measures for both studies in~\cref{tab:NASATLX-partA-transposed}.
Additionally, we provide the UEQ-S benchmark graphs in~\cref{fig: UEQ benchmark graphs}. 
\FloatBarrier
\begin{table}[H]
\caption{SSQ scores for study 1 (left) and study 2 (right).}
\resizebox{\columnwidth}{!}{%
\begin{tabular}{rcccc}
\toprule
     & \textbf{Nausea} & \textbf{Oculomotor} & \textbf{Disorientation} & \textbf{Total SSQ} \\
\midrule
\textbf{Prestudy Baseline} & $8.94\pm 10.1$ & $10.4\pm 14.9$ & $6.09\pm 21.4$ & $95.2\pm 125$ \\
\textbf{Study Block 1} & $12.5\pm 13.8$ & $23.2\pm 21.2$ & $28.7\pm 24.1$ & $241.0\pm 198.0$ \\
\textbf{Study Block 2} & $13.1\pm 10.4$ & $33.6\pm 21.1$ & $33.1\pm 29.1$ & $299.0\pm 205.0$ \\
\hline
\textbf{Prestudy Baseline} & $9.50\pm 12.2$ & $20.1\pm 21.8$ & $13.1\pm 27.2$ & $160.0\pm 214.0$ \\
\textbf{Study Block 1} & $20.2\pm 27.4$ & $29.9\pm 38.4$ & $29.5\pm 58.6$ & $298.0\pm 453.0$ \\
\textbf{Study Block 2} & $22.4\pm 31.4$ & $38.8\pm 42.1$ & $33.6\pm 63.6$ & $355.0\pm 495.0$ \\
\textbf{Study Block 3} & $23.0\pm 34.4$ & $38.8\pm 43.0$ & $35.2\pm 68.0$ & $363.0\pm 521.0$ \\
\bottomrule
\end{tabular}
}
\label{tab:ssq-transposed}
\end{table}

\begin{table}[H]
\caption[]{NASA RAW TLX measurements for each tested interaction technique. Left: Scores study 1, right: scores study 2.}
\centering
\resizebox{\columnwidth}{!}{%
\begin{tabular}{rll|lll}
\toprule 
       & \textbf{Gaze+Pinch} & \textbf{Gaze+Blink} & \textbf{Gaze+Pinch} & \textbf{Gaze+Blink} & \textbf{Gaze+BlinkP} \\
\midrule
\textbf{Mental Demand} & $25.9\pm 22.7$ & $28.8\pm 24.1$
                       & $44.7\pm 29.2$ & $41.8\pm 26.0$ & $50.0\pm 29.3$\\
\textbf{Physical Demand} & $24.4\pm 28.0$ & $26.9\pm 25.0$
                         & $37.4\pm 26.3$ & $40.3\pm 28.1$ & $49.1\pm 30.4$\\
\textbf{Temporal Demand} & $23.4\pm 21.5$ & $21.2\pm 23.6$
                         & $42.9\pm 34.0$ & $35.6\pm 32.8$ & $42.6\pm 34.6$\\
\textbf{Performance} & $42.2\pm 25.0$ & $33.1\pm 21.6$
                     & $30.3\pm 22.1$ & $27.9\pm 26.0$ & $31.8\pm 25.9$\\
\textbf{Effort} & $37.2\pm 27.4$ & $40.0\pm 24.3$
                & $54.7\pm 30.9$ & $50.0\pm 30.9$ & $62.6\pm 25.0$\\
\textbf{Frustration} & $31.9\pm 27.5$ & $38.4\pm 31.6$
                     & $48.8\pm 36.5$ & $38.2\pm 30.2$ & $49.7\pm 32.1$\\
\textbf{Total Workload} & $31.9\pm 20.0$ & $30.3\pm 18.5$
                        & $41.1\pm 23.2$ & $39.0\pm 21.8$ & $47.6\pm 20.8$\\
\bottomrule
\end{tabular}%
}
\label{tab:NASATLX-partA-transposed}
\end{table}

\FloatBarrier

%% file: tables/features.tex
\newcolumntype{Y}{>{\centering\arraybackslash}X} 
\newcolumntype{L}{>{\raggedright\arraybackslash}X} 

\begin{tabularx}{\columnwidth}{>{}p{0.30\columnwidth}|L} 
    \toprule
    \textbf{Feature Name} & \textbf{Description} \\
    \midrule
    left pupil diameter  & Diameter in mm of the left pupil \\
    right pupil diameter & Diameter in mm of the right pupil \\ \hline
    left openness        & Openness of the left eye, expressed from 0 (=fully closed) to 1 (=fully open)\\
    right openness       & Openness of the right eye, expressed from 0 (=fully closed) to 1 (=fully open)\\ \hline
    left direction x     & X-component of the Gaze direction of the left eye\\
    left direction y     & Y-component of the Gaze direction of the left eye\\
    left direction z     & Z-component of the Gaze direction of the left eye\\ \hline
    right direction x    & X-component of the Gaze direction of the right eye\\
    right direction y    & Y-component of the Gaze direction of the right eye\\
    right direction z    & Z-component of the Gaze direction of the right eye\\
    \bottomrule
\end{tabularx}%

%% file: tables/architecture.tex
\begin{tabular}{c|c|c|c}
    \toprule
    layer name & Module type & input dim. & output dim.  \\
    \midrule
    layer\_0   & Linear Layer    & 50000 & 128\\ 
    norm\_0    & Batch Norm      & 128   & 128\\
    act\_0     & Mish Activation & 128   & 128\\
    \hline
    block\_a1   & ResNet Module b & 128   & 128\\
    block\_a2   & ResNet Module b & 128   & 128\\
    block\_a3   & ResNet Module a & 128   &  64\\
    \hline
    block\_b1   & ResNet Module b &  64   &  64\\
    block\_b2   & ResNet Module b &  64   &  64\\
    block\_b3   & ResNet Module a &  64   &  32\\
    \hline
    block\_c1   & ResNet Module b &  64   &  64\\
    block\_c2   & ResNet Module b &  64   &  64\\
    block\_c3   & ResNet Module a &  64   &  32\\
    \hline
    block\_c1   & ResNet Module b &  32   &  32\\
    block\_c2   & ResNet Module b &  32   &  32\\
    block\_c3   & ResNet Module b &  32   &  32\\
    \hline
    layer\_1    & Linear layer    & 32 & 2\\
    act\_1  & Softmax Activation  & 2  & 2\\
    \bottomrule
\end{tabular}

%% file: tables/allpassword.tex
\begin{tabular}{c|c|c}
    \toprule
    Password & Shifts & Distance on keyboard\\
    \midrule
    SbJYNCPXsMJgsOkrnN  &  8 Shifts & 108.0 \\
    fUgNKeqNDRKexaJtlO  & 10 Shifts & 116.4 \\
    jJLJvEiUELduQNiHvM  & 12 Shifts & 147.8 \\
    pPtSGYNFGQqETTycUn  &  9 Shifts & 106.6 \\
    xLiielBBSTYerMmyNE  &  7 Shifts &  95.4 \\
    zRCrRaBfdLRAsxGtJQ  & 12 Shifts & 103.7 \\
    FyxVdHddohOWaRABLi  &  9 Shifts &  96.3 \\
    ayMkEWPsAsjUGXtIvc  & 11 Shifts & 111.2 \\
    wIRYQKQoyvvSptyNwi  &  7 Shifts & 101.4 \\
    vZPUxCDNUZeWDLwXUa  &  9 Shifts &  89.3 \\
    eotJClgdGMtIWSyvcF  &  8 Shifts &  95.9 \\
    ZSrQbNglMfJcDNTguw  & 12 Shifts & 123.8 \\
    BHHzMaKwKsXsguITcS  & 12 Shifts & 107.6 \\
    IMljNncDxOZPIsmBbb  &  9 Shifts & 118.4 \\
    NPyawMcaMZRiiwlSNT  &  6 Shifts &  93.2 \\
    \bottomrule
\end{tabular}




%% file: tables/passwords.tex
\begin{tabular}{r|l|l}
    Password & Shifts & Edit distance on keyboard\\
    \toprule
    DrrqdZWdfXfartzTse  &  7 Shifts &  63.82 \\
    wUKFKGAdrQNgcXRyjt  &  7 Shifts &  74.51 \\
    YHJTyyOnGGfGfbiici  &  7 Shifts &  82.31 \\
    gkSiEnVxRiRmtgMFfC  & 12 Shifts & 129.56 \\
    rhshuLBIPNkepdmvol  &  3 Shifts &  87.05 \\
    \bottomrule
\end{tabular}